\documentclass[fleqn,10pt]{wlscirep}
\usepackage[utf8]{inputenc}
\usepackage[T1]{fontenc}
\usepackage{amsmath,amsfonts,amssymb}
\usepackage{lmodern}
\usepackage{mathtools}
\usepackage{commath}
\usepackage{braket}
\usepackage{array}
\usepackage{xcolor}
\usepackage{lineno}
\usepackage{multirow}
\usepackage[ruled, vlined]{algorithm2e}
\newcommand{\order}{\mathcal{O}}

\title{Tensor-structured algorithm for reduced-order scaling large-scale Kohn-Sham density functional theory calculations}

\author[1]{Chih-Chuen (Ian) Lin}
\author[1,2]{Phani Motamarri}
\author[1,3,*]{Vikram Gavini}
\affil[1]{Department of Mechanical Engineering, University of Michigan, MI 48109-2125, United States}
\affil[2]{Department of Computational and Data Sciences, Indian Institute of Science, Bangalore 560012, India}
\affil[3]{Department of Materials Science \& Engineering, University of Michigan, MI 48109-2125, United States}

\affil[*]{vikramg@umich.edu}

\begin{abstract}
We present a tensor-structured algorithm for efficient large-scale DFT calculations by constructing a Tucker tensor basis that is adapted to the Kohn-Sham Hamiltonian and localized in real-space. The proposed approach uses an additive separable approximation to the Kohn-Sham Hamiltonian and an $L_1$ localization technique to generate the 1-D localized functions that constitute the Tucker tensor basis. Numerical results show that the resulting Tucker tensor basis exhibits exponential convergence in the ground-state energy with increasing Tucker rank. Further, the proposed tensor-structured algorithm demonstrated sub-quadratic scaling with system size for both systems with and without a gap, and involving many thousands of atoms. This reduced-order scaling has also resulted in the proposed approach outperforming plane-wave DFT implementation for systems beyond 2,000 electrons. 
\end{abstract}
\begin{document}

\flushbottom 
\maketitle
\thispagestyle{empty}

\section*{Introduction}

Density functional theory (DFT) has been the workhorse of ab-initio materials simulations for over three decades, providing many key insights into materials properties and materials behavior. In order to study ground-state properties, based on the Hohenberg-Kohn theorem~\cite{DFT_HK} and the Kohn-Sham formulation~\cite{DFT_KS}, DFT reduces the Schr\"{o}dinger equation in 3$N_e$ spatial coordinates ($N_e$ denoting the number of electrons) to an equivalent problem in the electron-density that only depends on three spatial coordinates. This reduces the exponential computational complexity (with system-size) of solving the Schr\"{o}dinger equation to the cubic computational complexity of DFT. While DFT has enabled wide-ranging ab-initio calculations, with $\sim$ 1/4th of the computational resources on some public supercomputers utilized for DFT calculations, the cubic computational complexity has limited routine DFT calculations to typical system-sizes involving a few hundred atoms. In an attempt to enable DFT calculations on large-scale systems that are critical to understanding many aspects of complex materials phenomena, many efforts over the past three decades have focused on developing reduced-order scaling algorithms for electronic structure calculations~\cite{Linear_Scaling_review, Bowler2012, Vanderbilt1993, Car1993, GoedeckerColombo1994, Galli1995, Stephan1998, ONETEP, PhysRevB.79.115110, Lin2013, Motamarri2014}.  These approaches have either relied on a localized representation of the single-electron wavefunctions (such as Wannier functions~\cite{Vanderbilt1993}) or the exponential decay of the density-matrix in real-space, and have been demonstrated to provide close to linear-scaling complexity for materials with a gap. However, they have not been widely successful for metallic systems (without a gap) either due to the errors resulting from realizing locality in the wavefunctions in real-space or due to the higher prefactors that make these approaches computationally more expensive than the traditional cubic-scaling algorithms for system sizes of interest. In this work, we present an alternative direction by using tensor-structured ideas to achieve systematically convergent and efficient DFT calculations that exhibit sub-quadratic scaling for systems with and without a gap over system-sizes spanning many thousands of atoms.    

This line of work is motivated from a study that revealed a low-rank representation for the electronic structure using Tucker and canonical tensor decomposition~\cite{KHOROMSKIJ20095749,HACKBUSCH2007697,doi:10.1137/080730408}. Another study based on \textit{a posteriori} analysis showed that the rank required to approximate the electron density is only weakly dependent on the system-size~\cite{BLESGEN20122551}. These studies have thereby prompted the development of a tensor-structured approach for DFT calculations~\cite{Motamarri2016a}, where a Tucker tensor basis adapted to the Kohn-sham Hamiltonian was employed to solve the Kohn-Sham equations.  Importantly, the rank of the Tucker tensor basis was only weakly dependent with system-size for materials systems with and without a gap, which revealed the potential for realizing a reduced-order scaling approach for DFT calculations. However, since the constructed Tucker tensor basis is global, the representation of the Kohn-Sham Hamiltonian in this basis was dense. An $L_2$ localization scheme that was explored to localize the Tucker tensor basis did not provide sufficient locality to exploit sparsity in the Kohn-Sham Hamiltonian. With increasing system-size, albeit the slow growth in the size of the Hamiltonian matrix due to the weak rank dependence, the approach became computationally prohibitive for large systems and limited the system-sizes to a few hundred atoms. 

\begin{figure}[!htb]
\centering
\includegraphics[width=\linewidth]{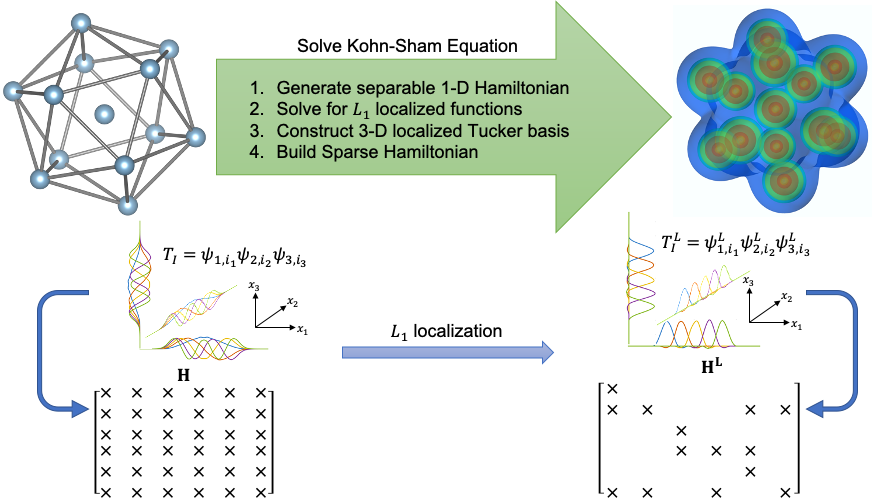}
\caption{Overview of the tensor-structured algorithm for Kohn-Sham DFT using $L_1$ localized functions. The tensor-structured algorithm seeks to construct a systematically convergent reduced-order tensor-structured basis for efficiently solving the Kohn-Sham equations. To this end, an additive separable approximation to the Kohn-Sham Hamiltonian is constructed, whose eigenbasis presents a suitable reduced order basis, given by $T_I=\psi_{1,i_1}\psi_{2,i_2}\psi_{3,i_3}$, where $I$ is a composite index $I = (i_1, i_2, i_3)$. However, the discrete Kohn-Sham Hamiltonian in this basis is dense due to the global nature of the 1-D functions $\psi_{k, i_k}$, $k = 1, 2, 3$ (lower-left). $L_1$ localization is applied to alleviate this bottleneck, where localized tensor-structured basis functions are constructed such that the subspace spanned by this localized basis is a close approximation to the eigensubspace of the separable Hamiltonian. Denoting the localized 1-D functions by $\psi^L_{k, i_k}$, the 3-D localized basis functions are given by $T^L_I=\psi^L_{1,i_1}\psi^L_{2,i_2}\psi^L_{3,i_3}$. The 3-D localized basis functions, being compactly supported, yields a sparse Hamiltonian (lower-right). The resulting sparse Hamiltonian, in conjunction with the slow growth of the Tucker rank with system-size to accurately represent the electronic structure, has provided sub-quadratic scaling with system-size for both insulating and metallic systems spanning over many thousands of atoms.}
\label{fig:schematic_plot}
\end{figure}

In this work, $L_1$ localization is used to overcome the aforementioned drawbacks, and we demonstrate systematically convergent, efficient and reduced-order scaling large-scale DFT calculations using tensor-structured techniques (cf. Fig.~\ref{fig:schematic_plot} for an overview of the approach). The $L_1$ localization utilizes the idea of auto-encoder commonly recognized in machine learning to construct a series of 1-D functions that are localized yet closely approximate the function space of interest. The 1-D localized functions that are a close approximation to the eigensubspace of a suitably constructed additive separable approximation of the Kohn-Sham Hamiltonian are used to generate the localized Tucker tensor basis for the DFT problem. The locality of the Tucker tensor basis results in a sparse discrete Kohn-Sham Hamiltonian matrix, which is exploited in the solution of the Kohn-Sham equations using the Chebyshev filtering subspace iteration scheme. The sparsity of the Kohn-Sham Hamiltonian matrix represented in the localized Tucker tensor basis improves both the computational efficiency and the memory footprint. Further, as will be demonstrated, the proposed approach has enabled sub-quadratic scaling DFT calculations on large-scale systems involving many thousands of atoms. The approach is generic and treats both systems with and without a gap on an equal footing. Importantly, this translates to substantial speed-ups over Quantum Espresso, a widely used state-of-the-art plane-wave DFT code~\cite{QE-2009,QE-2017}, with speed-ups of $\sim 8$-fold for metallic nano-particles containing $\sim 2,000$ atoms.  

\section*{Results}
The ground state energy in Kohn-Sham DFT (spin independent formulation) of a materials system with $N_a$ atoms and $N_e$ electrons is computed by solving a non-interacting single-particle Schrödinger equation in a mean field determined by the effective potential $V_\mathrm{eff}(\mathbf{x})$:
\begin{equation}
	\left( -\frac{1}{2}\nabla^2+V_\mathrm{eff}(\mathbf{x}) \right) \Psi_i = \epsilon_i \Psi_i, \quad i \in \{1, ..., N\}
    \label{eqn:ks eqn}.
\end{equation}
Equation~\ref{eqn:ks eqn} represents a non-linear eigenvalue problem with $\mathcal{H}:= -\frac{1}{2}\nabla^2+V_\mathrm{eff}$ being the Kohn-Sham Hamiltonian, and $\epsilon_i$ denoting the i-th Kohn-Sham eigenvalue and $\Psi_i$ denoting the corresponding Kohn-Sham orbital (eigenvector). The effective potential $V_\mathrm{eff}$ can be represented efficiently using a low-rank Tucker tensor approximation. We refer to the Supplementary Information (SI) for details on using tensor-structured techniques for an efficient computation of the effective potential, and subsequently the Hamiltonian matrix elements. The electron density $\rho(\mathbf{x})$ is computed in terms of the Kohn-Sham eigenstates as $\rho(\mathbf{x}) = 2\sum_{i=1}^{N}f(\epsilon_i; \mu)\left|\Psi_i(\mathbf{x})\right|^2$, where $f(\epsilon_i; \mu)$ denotes the orbital occupancy factor given by the Fermi-Dirac distribution $f(\epsilon; \mu) = 1/\left(1+exp(\frac{\epsilon - \mu}{k_B T})\right)$ with the Boltzmann constant $k_B$, the Fermi energy $\mu$, and the smearing temperature $T$. We note that Eq.~\ref{eqn:ks eqn} represents a non-linear eigenvalue problem, as the Kohn-Sham Hamiltonian depends on $\rho$, which in turn depends on the eigenstates. Thus, Eq.~\ref{eqn:ks eqn} is solved self-consistently via a self-consistent field (SCF) iteration\cite{martin_2004}.

\subsection*{Tensor-structured algorithm for Kohn-Sham DFT using $L_1$ localized 1-D functions}
In our previous work~\cite{Motamarri2016a}, it was suggested that an additive separable approximation to the Kohn-Sham Hamiltonian can be used to construct a Tucker tensor basis that is systematically convergent. In particular, using a tensor-structured cuboidal domain $\Omega$ spanned by tensor product of 1-D domains $\omega_{k=1,2,3}$, an additive separable approximation to the Kohn-Sham Hamiltonian ($\mathcal{H}_{1} (x_1) + \mathcal{H}_{2} (x_2) + \mathcal{H}_{3} (x_3) \approx \mathcal{H} (\mathbf{x})$ )  retains some features of the Hamiltonian, and thus presents a useful operator to generate reduced-order basis functions. 
To this end, the eigenfunctions of the additive separable approximation to the Hamiltonian, which constitute a Tucker tensor basis formed from the 1-D eigenfunctions of the separable parts of the Hamiltonian ($\mathcal{H}_k$, $k=1,2,3$), are used to solve the Kohn-Sham equations. While an efficient basis, the global nature of the ensuing Tucker tensor basis limits the computational efficiency of the algorithms to solve the Kohn-Sham equations. In the proposed work, in place of the 1-D eigenfunctions of $\mathcal{H}_k$, we instead construct compressed modes preserving the subspace spanned by the 1-D eigenfunctions using $L_1$ localization technique~\cite{Compressed_mode_osher}. The obtained 1-D localized functions are then used to generate the 3-D Tucker tensor basis, which is localized in real-space and allows us to exploit the sparsity of the Kohn-Sham Hamiltonian represented in this basis for both computational efficiency and realizing reduced-order scaling in solving the Kohn-Sham equations. The various aspects of our tensor-structured algorithm are now presented, which includes the generation of the additive separable approximation of the Kohn-Sham Hamiltonian, the evaluation of the $L_1$ localized 1-D functions, the construction of the localized Tucker tensor basis, the projection of the Kohn-Sham problem onto the localized Tucker tensor basis, and the solution of the Kohn-Sham equations. 

\subsubsection*{Construction of separable Hamiltonian}
We seek to construct a separable approximation to the Kohn-Sham Hamiltonian $\mathcal{H}_{1} (x_1) + \mathcal{H}_{2} (x_2) + \mathcal{H}_{3} (x_3) \approx \mathcal{H} (\mathbf{x})$ based on a rank-1 approximation of the eigenfunction corresponding to the lowest eigenvalue. 
To this end, we consider the rank-1 representation for the eigenfunction as $\Psi' (\mathbf{x})  = \psi_{1} (x_1)\psi_{2} (x_2)\psi_{3} (x_3)$. Thus, the problem of computing the smallest eigenvalue of the Kohn-Sham Hamiltonian using the rank-1 approximation is given by the variational problem 
\begin{equation}\label{eqn:variational_H}
    \min_{\psi_{k}}\, L(\Psi')\quad\quad \mbox{subject to:} \braket{\Psi'|\Psi'} = 1
\end{equation}
with the Lagrangian $L(\Psi') = \Bra{\Psi'} -\frac{1}{2}\nabla^2+V_\mathrm{eff}(\mathbf{x}) \Ket{\Psi'}$. Upon taking the variations of the functional with respect to $\psi_1$, $\psi_2$ and $\psi_3$, we obtain three simultaneous 1-D eigenvalue problems
\begin{equation}
        \mathcal{H}_k \psi_k = \alpha_k \psi_k, \quad k = 1, 2, 3.
    \label{eqn:1dks}
\end{equation}
As $\mathcal{H}_k$ and $\alpha_k$ are parametrized by $\psi_{l \neq k}$ (see SI for details), the three simultaneous 1-D eigenvalue problems represent a non-linear problem that can be solved self-consistently via SCF iteration. Upon achieving self-consistency, the 1-D Hamiltonians ($\mathcal{H}_k$) we obtain represent the additive separable approximation of the Kohn-Sham Hamiltonian that we seek. The eigenfunctions of this additive separable approximation to the Hamiltonian, which can be obtained as the tensor product of the 1-D eigenfunctions of $\mathcal{H}_k$ ($k=1,2,3$), constitute a complete basis, thus providing systematic convergence as will be demonstrated subsequently. 

 We note that the proposed approach represents one possibility of systematically constructing an additive separable approximation to the Kohn-Sham Hamiltonian, and other possibilities may exist. We also note that the resulting tensor-structured basis---the eigenbasis of the $\mathcal{H}_1+\mathcal{H}_2+\mathcal{H}_3$---is expected to be better than the plane-wave basis. To elaborate, the plane-wave basis is the eigenbasis of the Laplace operator (which is additive separable), whereas the additive separable approximation obtained via the proposed approach includes both the Laplace operator and an additive separable approximation of the Kohn-Sham potential $V_\mathrm{eff}$, thus retaining some additional features of the Kohn-Sham Hamiltonian and providing a better basis than the plane-wave basis. The superior approximation properties of the proposed tensor-structured basis over the plane-wave basis will be demonstrated subsequently via numerical benchmark studies (cf. Table.~\ref{tab:full_data}).

 We further note that the Kohn-Sham Hamiltonian changes during the course of the SCF iteration. In principle, the separable approximation to the Kohn-Sham Hamiltonian can be computed for each SCF iteration adapting the tensor-structured basis to the Hamiltonian in a given SCF iteration. However, in this work, we choose to keep the basis fixed after computing the tensor-structured basis in the first SCF iteration using the Kohn-Sham potential obtained from the superposition of atomic densities. This is motivated from our numerical studies on the benchmark examples involving both metallic and insulating systems, which demonstrate that the approximation properties of the resulting tensor-structured basis are not substantially altered. In particular, the difference in the ground-state energies obtained by either using a fixed basis constructed in the first SCF or adapting the basis in every SCF is substantially smaller than the basis discretization error. We refer to SI (section 6) for data supporting this observation. 
 
\subsubsection*{Computation of the $L_1$ localized 1-D functions} 
The tensor-structured basis computed using the 1-D eigenfunctions of $\mathcal{H}_k$ represents an efficient basis. However, the global nature of the basis limits the computational efficiency and scaling (with system size) of solution to the Kohn-Sham equations. To this end, we use an $L_1$ localization approach~\cite{Compressed_mode_osher} to construct a spatially localized tensor-structured basis that is a close approximation to the original tensor-structured basis. The localized basis is obtained by solving the following variational problem (for $k=1,2,3$)
\begin{equation}
     \min_{\mathbf{\Psi}'_k \in \mathbb{R}^{n \times N_k}} \frac{1}{\mu} \abs{\mathbf{\Psi}'_k} + \mathrm{Tr}({\mathbf{\Psi}'_k}^T \mathbf{H}_k \mathbf{\Psi}'_k) \quad \textrm{s.t.}\,\, {\mathbf{\Psi}'_k}^T\mathbf{\Psi}'_k=I,
    \label{eqn:L1_constraint}
\end{equation}
where $\mathbf{H}_k$ is matrix representation of $\mathcal{H}_k$ in a suitable orthogonal basis with dimension $n$, $\mathbf{\Psi}'_k$ denotes the representation of $N_k$ trial localized functions in the chosen basis, and $\mu$ is a parameter controlling the trade-off between the representability of the original eigensubspace and the locality of the 1-D functions, with $\abs{\cdot}$ denoting the $L_1$ norm of the matrix. The minimizer of this variational problem, henceforth denoted as $\mathbf{\Psi}^{L}_{k}$, provides localized functions whose span closely approximates the eigensubspace of the lowest $N_k$ eigenfunctions of $\mathcal{H}_k$, as will be demonstrated subsequently. We refer to the methods section for the solution procedure employed to solve the aforementioned variational problem. 
 
 \subsubsection*{Construction of the localized 3-D Tucker tensor basis $\mathbb{T}^L$}
The 1-D localized functions whose span is a close approximation to the subspace spanned by the 1-D eigenfunctions of $\mathcal{H}_k$ are subsequently used to construct the 3-D Tucker tensor basis. Denoting the 1-D localized functions as $\psi_{1, i_1}^L(x_1)$, $\psi_{2, i_2}^L(x_2)$, $\psi_{3, i_3}^L(x_3)$, the 3-D localized tensor-structured basis functions $T^L_I$ are given by
\begin{equation}
	T^L_I = \psi_{1, i_1}^L(x_1) \psi_{2, i_2}^L(x_2) \psi_{3, i_3}^L(x_3)
	\label{3-D basis},
\end{equation}
where $1 \leq i_d \leq R_d$ and $I$ is the composite index $I = (i_1, i_2, i_3)_{1 \leq i_d \leq R_d}$. The rank of the Tucker tensor basis is given by $(R_1, R_2, R_3)$ which denotes the number of localized 1-D functions in each direction. The space spanned by the 3-D localized tensor-structured basis functions is denoted as $\mathbb{T}^L$.

\subsubsection*{Projection of the Kohn-Sham Hamiltonian onto $\mathbb{T}^L$}
 The Kohn-Sham Hamiltonian is projected onto $\mathbb{T}^L$ spanned by the 3-D tensor-structured localized basis functions. We note that the Kohn-Sham effective potential $V_{\mathrm{eff}}$ is a functional of the electron-density $\rho$, and is comprised of a local-part $V_{\mathrm{eff}}^{loc}$ (local in real-space) and a non-local part $V_{\mathrm{ext}}^{nl}$. $V_{\mathrm{eff}}^{loc}$ includes the Hartree potential ($V_\text{H}$), the exchange-correlation potential and the local-part of the pseudopotential, whereas $V_{\mathrm{ext}}^{nl}$ comprises of the non-local projectors of the pseudopotential (cf. SI section 1). The convolution integral involved in the evaluation of $V_\text{H}$ can be efficiently computed using a low-rank Tucker tensor decomposition of the electron density ($R_{\rho}$ denoting the rank of the decomposition) and approximating the coulomb integral by a series of Gaussian functions~\cite{KHOROMSKIJ20095749, Hackbusch_kernel_approx} (cf. SI section~1.1 for details). Subsequently, a low-rank Tucker tensor approximation of $V_{\mathrm{eff}}^{loc}$ and $V_{\mathrm{ext}}^{nl}$ is utilized, with $R_V$ and $R^{nl}_V$ denoting the corresponding ranks, respectively. Denoting the low-rank Tucker approximation of the effective potential $V_{\mathrm{eff}}$ by $\tilde{V}_{\mathrm{eff}}$ (cf. SI section~1 for details), whose approximation error decays exponentially with the Tucker rank~\cite{KHOROMSKIJ20095749,HACKBUSCH2007697}, the projection of the Kohn-Sham Hamiltonian onto $\mathbb{T}^L$ is given by 
\begin{equation}
	\tilde{H}^L_{I, J} = \Bra{T^L_{I}}  -\frac{1}{2}\nabla^2+\tilde{V}_\mathrm{eff}(\rho; \mathbf{R}) \Ket{T^L_{J}}\,.
	\label{KS-Hamiltonian projection}
\end{equation}
 We note that the low-rank representation $\tilde{V}_{\mathrm{eff}}$ reduces the integrals involved in Eq.~\eqref{KS-Hamiltonian projection} to tensor products of one-dimensional integrals, thereby facilitating efficient evaluation of Hamiltonian matrix elements (cf. SI section 4).

The projected Kohn-Sham Hamiltonian matrix elements are computed, and a truncation tolerance is introduced to zero out the Hamiltonian matrix elements below the tolerance. This truncation is performed in every SCF iteration, which improves the sparsity of the Hamiltonian matrix and thereby reducing the memory footprint of the calculation. Furthermore, the sparsity of the Hamiltonian matrix also reduces the computational complexity of the algorithm employed to solve the Kohn-Sham equations, which is discussed subsequently. We note that the error in the ground-state energy, corresponding to the truncation of the Hamiltonian matrix elements, systematically decreases with tighter truncation tolerance. We refer to SI (section 7, Table~SI2) which provides data to support this observation. Furthermore, we note that a truncation tolerance of 1e-4 provides excellent sparsity in the Hamiltonian, with the error resulting from the truncation being significantly smaller than the basis discretization error and the desired chemical accuracy in ground-state energy (cf. SI section 7).

\subsubsection*{Computation of the occupied eigenstates}
The discretized Kohn-Sham problem, corresponding to Eq.~\ref{eqn:ks eqn}, in the localized orthonormal tensor-structured basis is given by the standard eigenvalue problem
\begin{equation}
	\mathbf{H}^{L}\mathbf{\Psi}_i = \epsilon_i\mathbf{\Psi}_i \,, \quad i\in\{1,\ldots,N\}
	\label{eqn:localized ks problem}
\end{equation}
where $\mathbf{H}^{L}$ denotes the truncated sparse Kohn-Sham Hamiltonian matrix. We use the Chebyshev filtering based subspace iteration (ChFSI)~\cite{ChFSIoriginal} to efficiently solve the Kohn-Sham equations.  The ChFSI method has been demonstrated to be efficient with good parallel scalability for real-space implementations of DFT~\cite{Motamarri2013b,DFT-FE}. In the ChFSI method, in each SCF iteration, a suitably constructed Chebyshev filter using $\mathbf{H}^{L}$ is employed to construct a close approximation to the relevant eigensubspace of the occupied states. The action of the Chebyshev filter on a given subspace can be cast as a recursive iteration involving matrix-vector multiplications between $\mathbf{H}^{L}$ and vectors obtained during the course of recursive iteration. Since $\mathbf{H}^{L}$ is sparse, the computational complexity of the Chebyshev filtering operation scales as $\order{(R^{3}N})$, where $R=\max\{R_1,R_2,R_3\}$. In ChFSI, the Chebyshev filtered vectors are orthogonalized using a Gram-Schmidt orthogonalization procedure, and subsequently the Kohn-Sham eigenstates are computed by projecting $\mathbf{H}^{L}$ onto the Chebyshev filtered subspace and diagonalizing this projected Hamiltonian. The computational complexity of the orthogonalization procedure and the subspace projection scales as $\order{(R^{3}N^2)}$ while the diagonalization cost scales as $\order{(N^3)}$. As demonstrated in Table~SI3 (SI section 8), Chebyshev filtering, which scales linearly with $N$, remains the dominant cost even at 25,000 electrons for the various benchmark examples considered in this work. We note that, at even larger system sizes, other costs that exhibit quadratic-scaling (orthogonalization and subspace projection) and cubic-scaling (diagonalization) with $N$ can start to compete. However, at such a point, explicit diagonalization can be avoided, and already developed ideas~\cite{Motamarri2014} of localizing the Chebyshev filtered vectors in conjunction with Fermi-operator expansion can be adopted to retain the reduced-order scaling for systems with or without a gap. We refer to the methods section for details on the various numerical parameters employed in conducting the DFT calculations using self-consistent field iteration approach via the ChFSI technique.
 
\subsection*{Eigensubspace representability of the localized 1-D functions}
\begin{figure}[!htb]
\centering
\includegraphics[width=\linewidth]{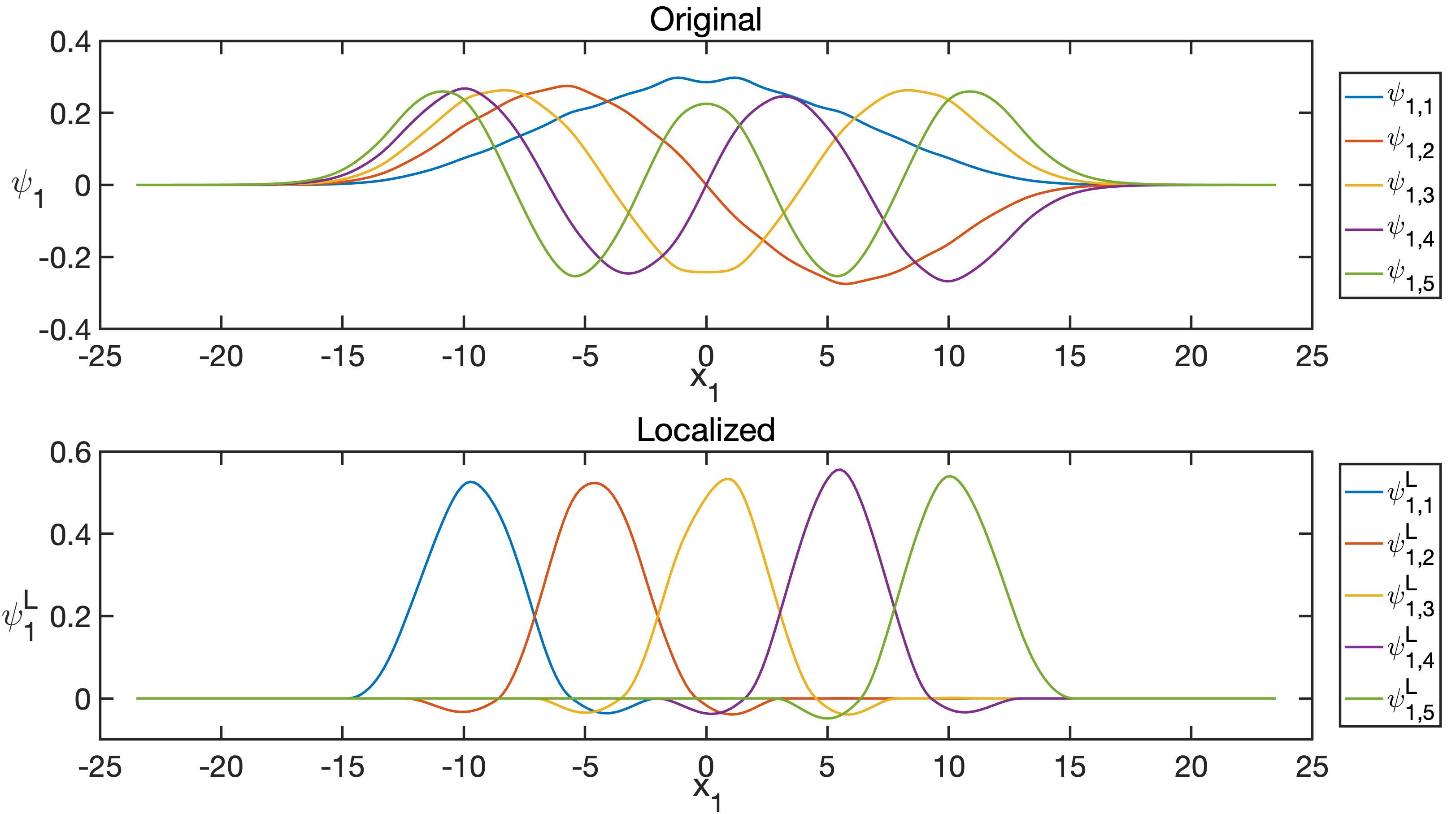}
\caption{1-D functions in $x_1$ direction constructed from the additive separable approximation of the Kohn-Sham Hamiltonian  for $\mathrm{Al}_{147}$ nano-particle. Top: Lowest five eigenfunctions of $\mathcal{H}_1$. Bottom: The corresponding $L_1$ localized 1-D functions.}
\label{fig:localized_basis_comparison}
\end{figure}

We now demonstrate the ability of the $L_1$ localized functions to closely approximate the eigensubspace of $\mathcal{H}_k$ using $\mathrm{Al}_{147}$ nano-particle with icosahedral symmetry. We compute the additive separable approximation of the Kohn-Sham Hamiltonian for this nano-particle, and, then compute the lowest 70 eigenstates of $\mathcal{H}_k$. We subsequently use $L_1$ localization approach to compute the localized functions that are a close approximation to the eigensubspace. Figure~\ref{fig:localized_basis_comparison} shows the lowest 5 eigenfunctions of $\mathcal{H}_1$ (one of the 1-D separable Hamiltonian) (top) and the corresponding 1-D localized functions (bottom). 
We refer to SI (Fig.~SI1) for an illustration of all 70 eigenstates and the corresponding 1-D localized functions. It is evident that, while the eigenfunctions are global in nature, the functions obtained from the $L_1$ localization approach are localized in real-space. This locality is key to the sparsity of the Kohn-Sham Hamiltonian matrix in the Tucker tensor basis, and the resulting computational efficiency.   

\begin{figure}[!htb]
\centering
\includegraphics[width=0.7\linewidth]{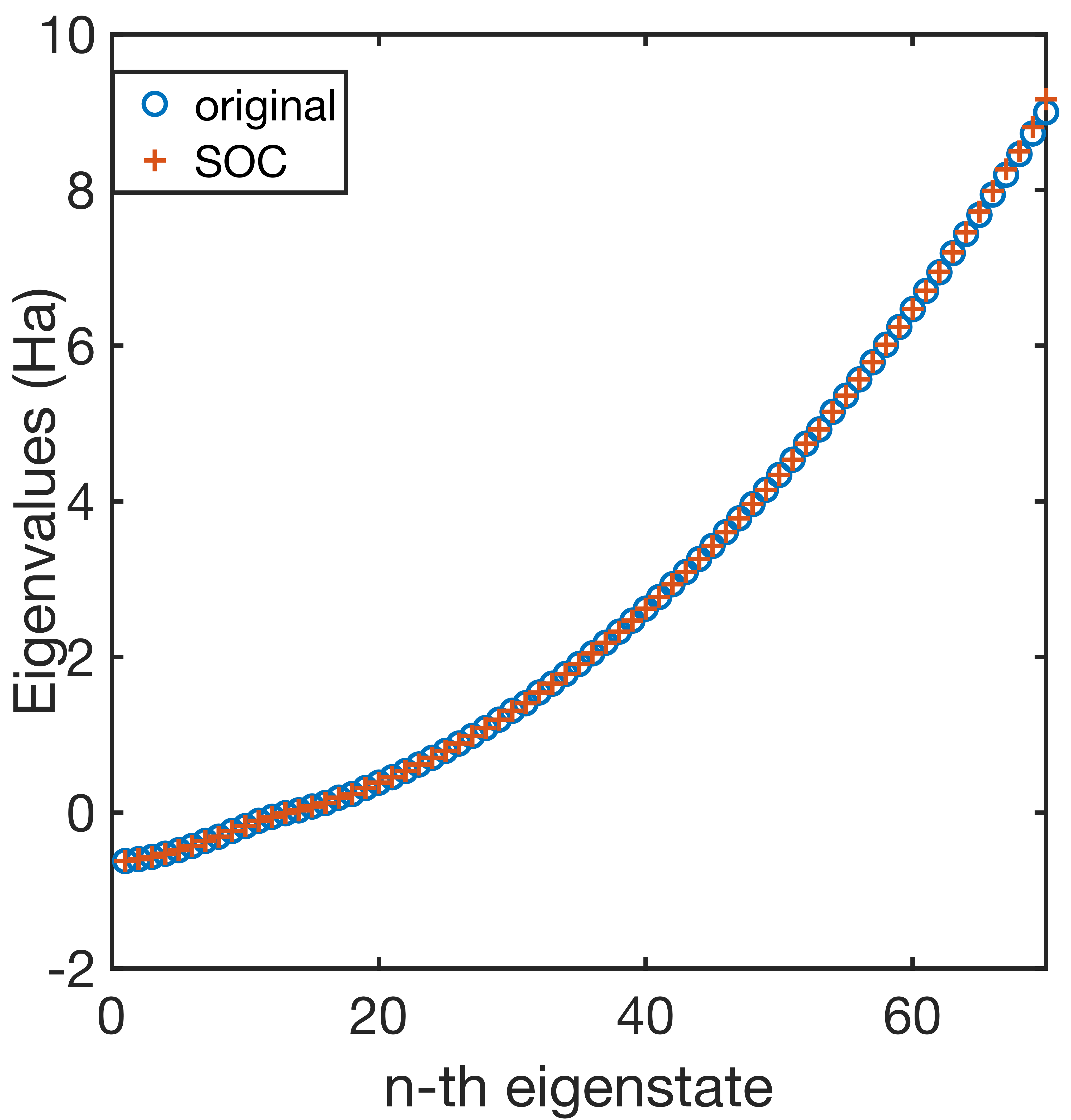}
\caption{Comparison of the eigenvalues of the 1-D separable Hamiltonian in $x_1$ direction of $\mathrm{Al}_{147}$ nano-particle (marked with blue circle) with the eigenvalues of $K_{ij}=\Bra{\psi^L_{1,i}} \mathcal{H}_{k} \Ket{\psi^L_{1,j}}$ (marked with red cross).}
\label{fig:shell3 error}
\end{figure}

In order to demonstrate the accuracy of the $L_1$ localization approach in closely approximating the eigensubspace of the separable Hamiltonian, we consider the first 70 eigenstates of $\mathcal{H}_1$ and the eigenvalues of the matrix $K_{ij}=\Bra{\psi^L_{1,i}} \mathcal{H}_1 \Ket{\psi^L_{1,j}}$, $1\leq i,j, \leq 70$. Figure~\ref{fig:shell3 error} shows the eigenvalues of $\mathcal{H}_1$ and the eigenvalues of $K_{ij}$. It is interesting to note that the eigenvalues of the first 65 states are almost identical, with only slight deviations for the higher states. This demonstrates that the space spanned by the localization functions obtained using the $L_1$ localization approach is a close approximation to the eigensubspace of $\mathcal{H}_k$. We also note here that better accuracy can be achieved, when necessary, by simply increasing the size of $N_k$ to be solved for in Eq.~\ref{eqn:L1_constraint}.  In order to assess the accuracy afforded by the localization procedure in the ground-state energy, we computed the ground-state energy of $\mathrm{Al}_{147}$ using the 3-D localized basis with rank 70, and compared that with the energy obtained using the eigenbasis of $\mathcal{H}_{1}+\mathcal{H}_{2}+\mathcal{H}_{3}$ (i.e., without localization) of the same rank. The energy obtained without localization is -56.61882 eV/atom in comparison to -56.61893 eV/atom obtained using localization. Thus, the error introduced due to localization is $\sim$0.1meV/atom, which is substantially smaller than the basis discretization error of $\sim$8.5 meV/atom corresponding to rank 70 (reference energy is -56.6274 eV/atom; cf. Table~\ref{tab:full_data}).

\subsection*{Convergence of the tensor-structured basis}
We next investigate the convergence properties of the 3-D Tucker tensor basis constructed from the 1-D localized functions. 
For the convergence study we consider two benchmark problems: (i) $\mathrm{C_{60}}$  (fullerene) molecule; (ii) tris (bipyridine) ruthenium, a transition metal complex. We note that these systems have no tensor structure symmetry and serve as stringent benchmarks to assess the convergence and accuracy afforded by the proposed Tucker tensor basis.  The ground-state energy for these molecules is computed for various Tucker tensor ranks $R$ ($R_1=R_2=R_3=R$), with the 3-D Tucker tensor basis getting systematically refined with increasing $R$. In this study, $R_{\rho}$, $R_{V}$ and $R^{nl}_{V}$ are chosen stringently such that the resulting errors are significantly smaller than the basis discretization errors, and they are held fixed for increasing $R$. In particular, we used $R_{\rho}=45$, $R_{V}=65$ and $R^{nl}_{V}=20$ for fullerene, and $R_{\rho}=80$, $R_{V}=55$ and $R^{nl}_{V}=20$ for tris (bipyridine) ruthenium. The basis discretization error (convergence with respect to $R$) is measured with respect to a well-converged Quantum Espresso result. The converged Quantum Espresso ground-state energies for fullerene molecule is taken to be -155.1248 eV/atom ($E_{cut} = 60$ Ha) and that of tris (bipyridine) ruthenium is taken to be -118.2128 eV/atom ($E_{cut} = 65$ Ha).

\begin{figure}[!htb]
    \centering
    \begin{tabular}{cc}
         \includegraphics[width=0.45\linewidth]{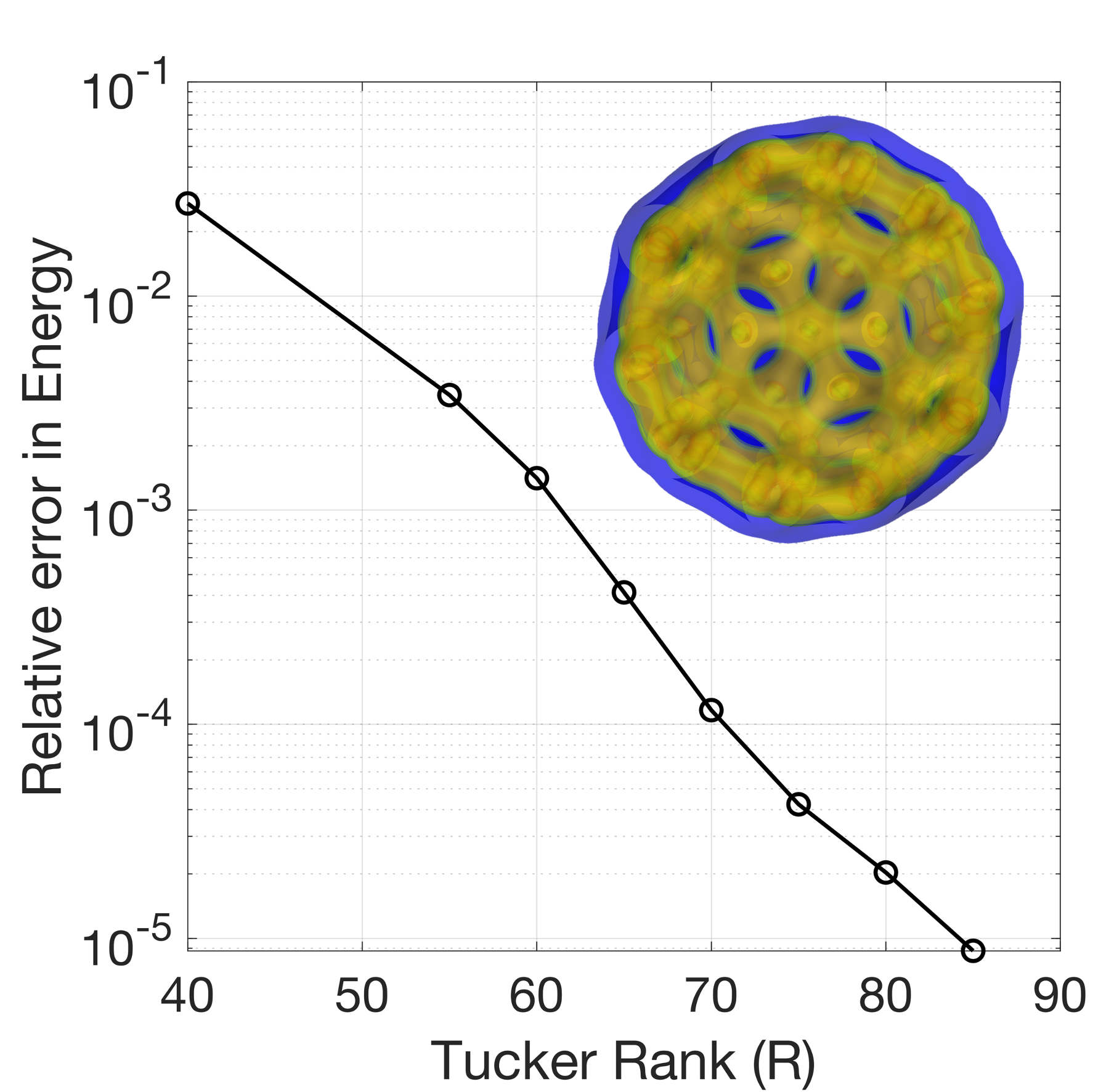} &  \includegraphics[width=0.45\linewidth]{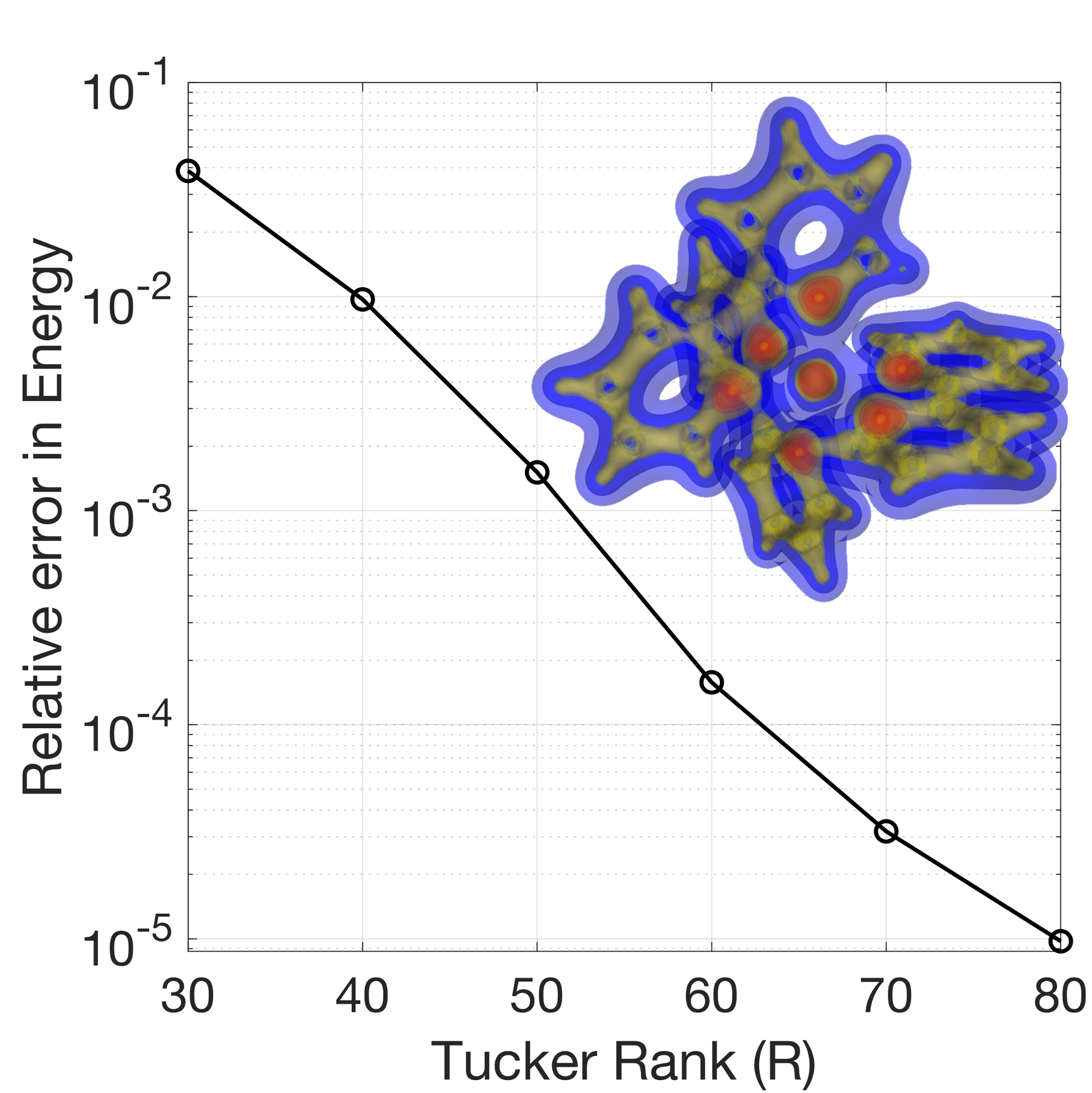}\\
         (a) & (b)
    \end{tabular}
    \caption{ Convergence with respect to the Tucker rank: (a) Fullerene; (b) Tris (bipyridine) ruthenium. The electron density isocontours are provided in the inset.}
    \label{fig:Convergence plot}
\end{figure}

 Figure~\ref{fig:Convergence plot}(a) and Fig.~\ref{fig:Convergence plot}(b)  show the relative error in the ground-state energy for the various ranks of the Tucker tensor basis. It is evident from these results that the Tucker tensor basis constructed using our approach provides an exponential convergence in the ground-state energy with increasing Tucker rank. The convergence study of these molecules suggests that the proposed tensor-structured technique provides systematic convergence with high accuracy and is capable of handling generic materials systems, including those involving transition metals. 

\subsection*{Performance and scaling analysis}
To study the performance and scaling with system-size of the proposed tensor-structured approach for DFT calculations, we consider two classes of benchmark systems: (i) Aluminum nano-particles of various sizes ranging from 13 atoms to 6,525 atoms; (ii) Silicon quantum dots with system sizes ranging from 26 atoms to 7,355 atoms. These benchmark systems constitute materials systems with and without a gap, thus allowing us to assess the system-size scaling for both classes of materials. 
In order to compare the efficiency of the proposed tensor-structured approach with the widely used plane-wave DFT calculations, we also conducted the DFT calculations using Quantum Espresso wherever possible. For the sake of estimating the computational efficiency, the energy cut-off for Quantum Espresso and the Tucker rank are chosen such that the ground-state energy is converged to within 10 meV/atom measured with respect to a highly converged reference calculation. The reference ground-state energies are obtained from Quantum Espresso (using a high energy cut-off) for smaller systems and the DFT-FE code~\cite{DFT-FE}---a massively parallel real-space code for large-scale DFT calculations---for larger system sizes. The cell size for plane-wave calculations is chosen such that each atom is at least 10 Bohr away from the boundary, which was needed to obtain the desired accuracy. 

 In these benchmark calculations, the additive separable approximation to the Kohn-Sham Hamiltonian is computed only in the first SCF iteration and the resulting 3-D Tucker basis is held fixed for subsequent SCF iterations. We note that the approximation properties of an adaptive Tucker basis (where the basis is regenerated for every SCF iteration) and the fixed Tucker basis are similar, with the differences in the accuracy being substantially smaller than the basis discretization error for a given Tucker rank. We refer to SI section 6 (cf. Table SI1) which provides data supporting this observation. In the tensor structured calculations reported subsequently, all the numerical parameters---ranks for approximating electron density and effective Kohn-Sham potential $(R_{\rho}, R_{V}, R^{nl}_{V})$ in $\tilde{\mathbf{H}}^{L}$, and the truncation tolerance adopted in computing $\mathbf{H}^{L}$---are chosen such that the resulting errors are substantially smaller than the basis discretization error in the ground-state energy associated with the Tucker rank $R$ of the localized 3-D Tucker tensor basis and the desired chemical accuracy. In particular, for the Aluminum nano-particles we used $R_{\rho}=40$, $R_V=50$ and $R^{nl}_{V}=25$. In the case of Silicon quantum dots, we used $R_{\rho}=55$, $R_V=55$ and $R^{nl}_{V}=25$. The truncation tolerance in computing $\mathbf{H}^{L}$ was chosen to be 1e-4 for all the calculations, which provides excellent sparsity for $\mathbf{H}^{L}$, and, importantly, the sparsity is either steady or improves with increasing system size (cf. Table~SI2 \& SI3). The error in ground-state energy associated with this choice of truncation tolerance is $\sim1$ meV/atom (cf. Table~SI2), as opposed to the targeted accuracy in this study of being within $10$ meV/atom of reference energies. 

\subsubsection*{Aluminum nano-particles}
\begin{figure}[!htb]
    \centering
    \begin{tabular}{cc}
         \includegraphics[width=0.5\linewidth]{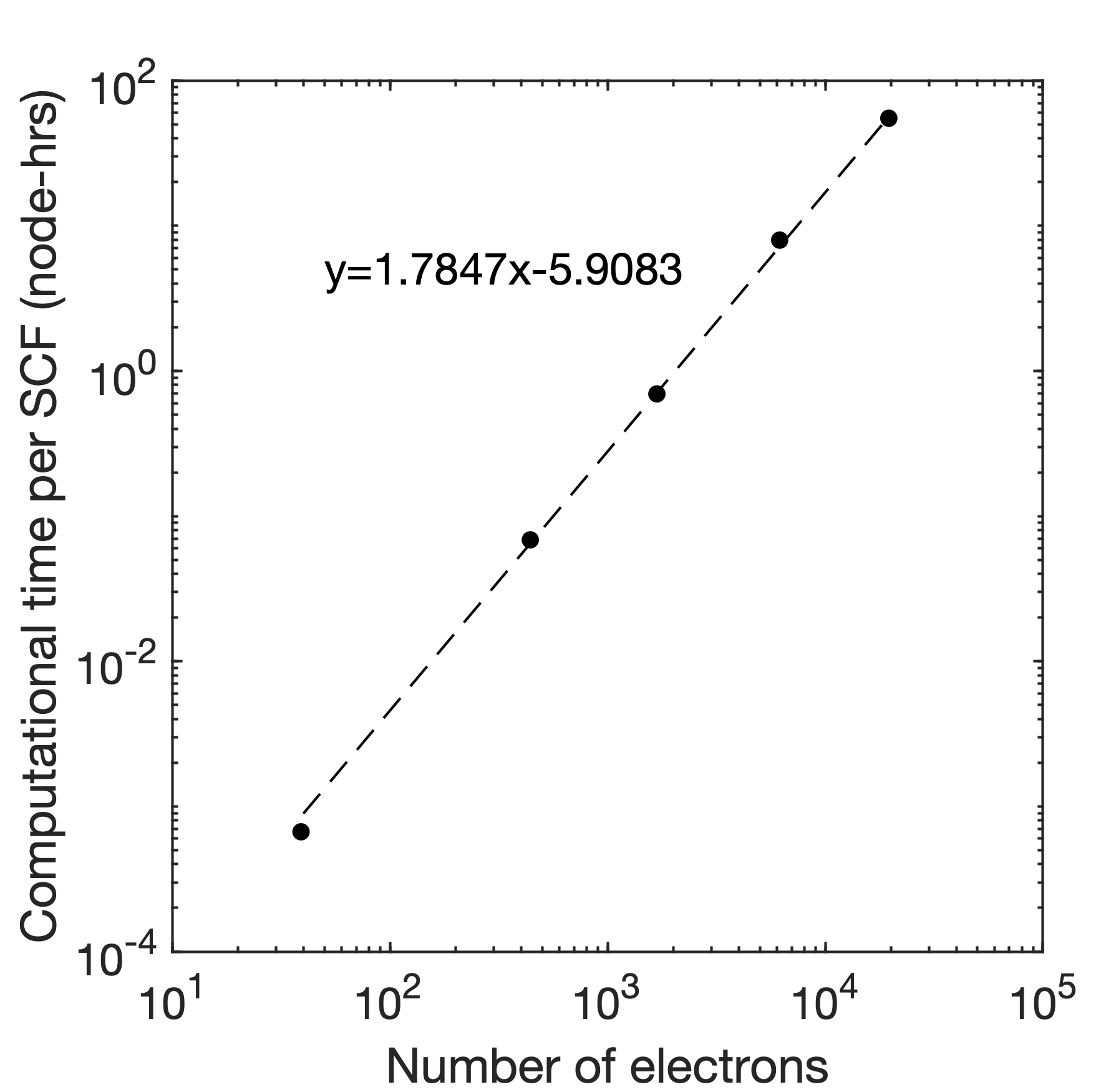} &  \includegraphics[width=0.5\linewidth]{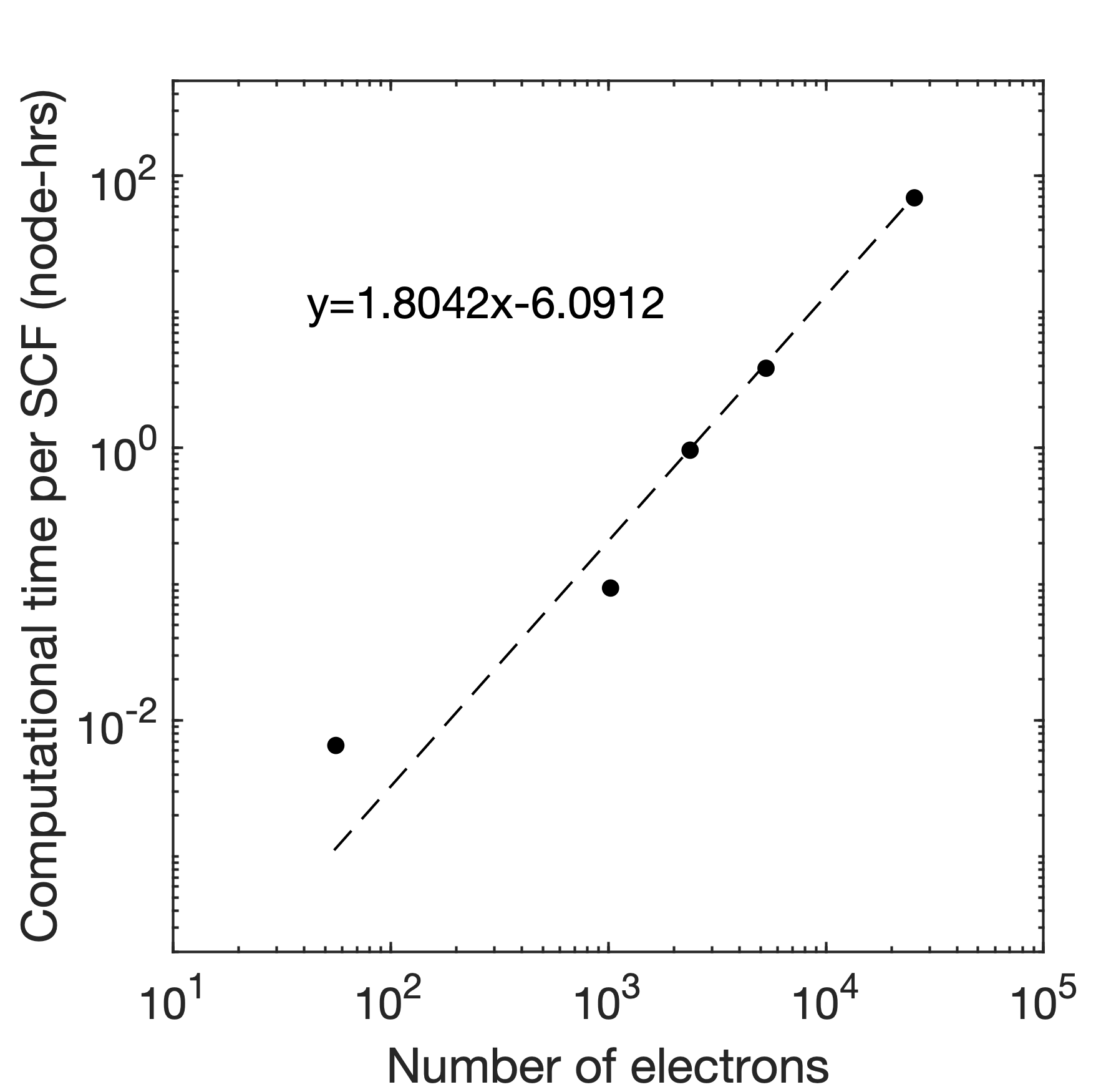}\\
         (a) & (b)
    \end{tabular}
    \caption{ Computational complexity of the tensor-structured approach: (a) Aluminum nano-particles $\order(1.78)$; (b) Silicon quantum dots $\order(1.8)$. }
    \label{fig:scaling plot}
\end{figure}

The  computational efficiency afforded by  the proposed tensor-structured approach  in comparison  to Quantum Espresso for the various aluminum nano-particles with icosahedral symmetry considered in this work are provided in  Table.~\ref{tab:full_data}(a) .  We note that the Tucker rank required to achieve the desired accuracy only grows slowly with increasing system-size. Importantly, we note that the number of basis functions needed to achieve the desired accuracy using the localized Tucker basis is smaller than the plane-wave basis. As previously discussed, this is a consequence of the superior approximation properties of the Tucker tensor basis generated as the eigenbasis of an additive separable approximation of the Kohn-Sham Hamiltonian that in addition to the Laplace operator retains some characteristics of the Kohn-Sham potential, as opposed to the plane-wave basis which corresponds to the eigenbasis of the Laplace operator.
In terms of the computational time, while Quantum Espresso is more efficient for the smaller system-sizes, the tensor-structured approach starts to substantially outperform for larger system sizes. Notably, for $\mathrm{Al}_{2057}$, the tensor-structured approach is 8-fold more efficient.
Furthermore, using the computational times, the scaling of the proposed tensor-structured approach is estimated to be around $\mathcal{O}(N_e^{1.78})$ with $N_e$ denoting the number of electrons  (cf. Fig.~\ref{fig:scaling plot}(a)). Notably, the scaling with system-size is sub-quadratic for this metallic system over system-sizes spanning many thousands of atoms, as opposed to the cubic-scaling complexity for plane-wave DFT calculations.  We note that this is a consequence of the slow growth of the Tucker rank with system-size that results in a sub-linear growth of the total number of basis functions with system-size. The breakdown of the computational costs for the various steps of the calculation is provided in SI section 8 (cf. Table SI3). 

\subsubsection*{Silicon quantum dots}
 Table~\ref{tab:full_data}(b)  compares the computational performance of the proposed tensor-structured approach with Quantum Espresso for a wide range of silicon quantum dots passivated with hydrogen. As in the case of aluminum nano-particles, the  Tucker tensor basis is more efficient than the plane-wave basis in terms of the number of basis functions to attain the desired accuracy. In terms of computational time, the proposed tensor-structured approach  starts competing with Quantum Espresso beyond a few hundred atoms, and significantly outperforms for larger systems. 
Moreover, the scaling with system size for the tensor-structured algorithm, for a range of system-sizes with the largest containing 7,355 atoms, is estimated to be $\mathcal{O}(N_e^{1.8})$  (cf. Fig.~\ref{fig:scaling plot}(b)). Notably, this scaling is similar to that obtained for aluminum nano-particles as the algorithm treats systems with and without a gap on a similar footing.

\section*{Discussion}

We have presented a tensor-structured algorithm, where the Tucker tensor basis is constructed as a tensor product of localized 1-D functions whose span closely approximates the eigensubspace of a suitably constructed additive separable approximation to the Kohn-Sham Hamiltonian. The resulting localized Tucker tensor basis, that is adapted to the Kohn-Sham Hamiltonian, provides a systematically convergent basis as evidenced by the exponential convergence of the ground-state energy with increasing Tucker rank. Our numerical studies on the computational performance suggest that the proposed approach exhibits sub-quadratic scaling (with system-size) over a wide range of system-sizes with the largest involving many thousands of atoms. Importantly, the sub-quadratic scaling is realized for both systems with and without a gap, as the algorithm treats both metallic and insulating systems on an equal footing. Further, comparing the computational efficiency of the proposed approach with Quantum Espresso, we observe significant outperformance for system-sizes beyond 5,000 electrons.       

We note that the sub-quadratic scaling is a consequence of the slow growth of the Tucker rank with system-size, with the resulting number of basis functions growing sub-linearly with system-size even for systems containing many thousands of atoms. By combining the proposed approach with reduced-order scaling techniques that exploit the locality of the wavefunctions in real-space, there is further room to reduce the scaling with system-size and is a useful future direction to pursue. Further, the proposed tensor-structured approach is amenable to GPU acceleration that can further substantially enhance the computational efficiency of the approach, and is currently being pursued.  We note that the benchmark systems presented here were restricted to non-periodic calculations as the proposed tensor-structured approach was implemented in a non-periodic setting as a first step of an ongoing effort. However, we remark that the ideas presented here are generic and can be extended to periodic calculations. 

\section*{Methods}
\subsection*{Tucker tensor representation}
Tucker tensor representation is a higher-order generalization of principal component analysis for a tensor~\cite{Tucker1963a,tucker64extension,Tucker1966mathnotes}. An N-way tensor is approximated by a Tucker tensor through Tucker decomposition with a smaller N-way core tensor and N factor matrices whose columns are the rank-1 components from the decomposition~\cite{Tucker1963a, tucker64extension, Tucker1966mathnotes, doi:10.1137/S0895479896305696,Kolda2009a}. In the scope of this work, the discussion is restricted to three-way tensor. Let $A \in \mathbb{R}^{I_1 \times I_2 \times I_3}$ be a real-valued three-way tensor of size $I_1 \times I_2 \times I_3$ indexed by a set of integers $(i_1, i_2, i_3)$ 
\begin{equation}
    A_{(i_1, i_2, i_3)} = a_{i_1i_2i_3},
    \label{eqn:tensor_entry_def}
\end{equation}
where $i_d \in \{1, 2, ..., I_d\}, I_d \in \mathbb{N}$ and $d \in \{1, 2, 3\}$ denotes the dimensions. A Tucker tensor representation of the tensor $A$ with decomposition rank $\mathbf{R}=(R_1, R_2, R_3)$ for each direction has the form
\begin{equation}
    A \approx A^{(\mathbf{R})} = \sum_{r_1=1}^{R_1} \sum_{r_2=1}^{R_2} \sum_{r_3=1}^{R_3} \sigma \mathbf{u}^
    {r_1}_{1} \circ \mathbf{u}^{r_2}_{2} \circ \mathbf{u}^{r_3}_{3},
    \label{eqn:ttensor_def}
\end{equation}
where $\sigma \in \mathbb{R}^{R_1 \times R_2 \times R_3}$ is the core tensor, $\mathbf{u}_d^{r_d} \in \mathbb{R}^{I_d}$ forms the factor matrix $\mathbf{U}_d \in \mathbb{R}^{I_d \times R_d}$, and "$\circ$" denotes the vector outer product $(\mathbf{u}^
    {r_1}_{1} \circ \mathbf{u}^{r_2}_{2} \circ \mathbf{u}^{r_3}_{3})_{i_1, i_2, i_3} := u^{r_1}_{1, i_1} u^{r_2}_{2, i_2} u^{r_3}_{3, i_3}$. The core tensor stores the coefficients $\sigma_{r_1 r_2 r_3}$ for each rank-1 tensor $\mathbf{u}^
    {r_1}_{1} \circ \mathbf{u}^{r_2}_{2} \circ \mathbf{u}^{r_3}_{3}$. The core tensor and the factor matrices can be viewed as the higher-order correspondence of the singular values and unitary matrices. 
The tensor representation of the Tucker form can be obtained with the higher-order singular value decomposition (HOSVD). The HOSVD flattens the given tensor in three directions and employs singular-value decomposition to obtain the factor matrices. The factor matrices are then used to contract with the given tensor to obtain the core tensor. We refer to \cite{Kolda2009a} and \cite{Hackbusch} for details of HOSVD and further review on tensor decomposition and tensor analysis. In this work, an MPI implementation in C++ for Tucker decomposition is used~\cite{AuBaKo16,BallardKK20}.

\subsection*{Computation of $L_1$ localized 1-D functions}
The variational problem in Eq.~\ref{eqn:L1_constraint} is solved by the splitting orthogonality constraint algorithm (SOC). The SOC algorithm introduces two auxiliary variables controlling the orthonormality and the locality constraints, and translates the variational problem into a constrained minimization problem. The constrained minimization problem is split into one minimization problem and two constraints. The SOC algorithm is capable of providing a set of compressed modes with good locality, yet preserving the orthogonality. We refer to Ozolinš et al.~\cite{Compressed_mode_osher} and Lai et al.~\cite{SOC} for a detailed discussion, and present the algorithm in the context of this work in the SI for the sake of completeness. 

\subsection*{Ground-state DFT calculations}
 All calculations are performed using the norm-conserving Troullier-Martin pseudopotentials in Kleinmann-Bylander form~\cite{PSP_TM, PSP_KB}, and a local density approximation (LDA) for the exchange-correlation functional~\cite{LDA_CA, LDA_PZ, LDA_PW}. The numerical parameters comprising the Tucker decomposition rank used in $\tilde{V}_{\mathrm{eff}}$ and the truncation tolerance used in the Hamiltonian matrix elements are chosen such that the numerical errors are lesser than the Tucker tensor basis discretization error and the desired chemical accuracy. In the ChFSI method employed to solve the Kohn-Sham equations, we use a Chebyshev filter constructed using polynomial degree of 10-20 for the various materials systems reported in this study. The n-stage Anderson mixing scheme \cite{anderson_mixing} is employed in the SCF iteration.   We used Fermi-Dirac smearing with $T=500$K for computing the orbital occupancies. The performance benchmarks are obtained on compute nodes comprising of 68-core Intel Xeon Phi Processor 7250 and 96GB memory per node.  All calculations were performed in the good parallel-scaling regime to ensure that the obtained computational times (node-hours) are representative of the computational efficiency of the approach.

\section*{Acknowledgements}
We gratefully acknowledge the support of the Air Force Office of Scientific Research through grant number FA-9550-17-0172 under the auspices of which this work was conducted. V.G. also gratefully acknowledges the support of the Army Research office through the DURIP grant W911NF1810242, which provided computational resources for this work.

\clearpage

\section*{Supplementary Information}
\renewcommand{\thesubsection}{S.\arabic{subsection}}
\renewcommand{\theequation}{SI\arabic{equation}}
\setcounter{equation}{0}
\section{Kohn-Sham effective potential computation using tensor-structured techniques}
Here we elaborate the various aspects of utilizing the low-rank tensor decomposition to evaluate and represent the various components of the Kohn-Sham effective potential. We express the effective potential $V_\mathrm{eff} (\mathbf{x})$ as $V_\mathrm{eff} (\rho; \mathbf{R}_{nu}$) to emphasize that the effective potential is a functional of the electron density $\rho$ and is parametrized by the coordinates of nuclei $\textbf{R}_{nu}=\{\mathbf{R}_1,\mathbf{R}_2, \ldots, \mathbf{R}_{N_a}\}$. The effective potential  can be decomposed as 
\begin{equation}
    V_\mathrm{eff}(\rho; \mathbf{R}_{nu}) = V^{loc}_{\mathrm{eff}}(\rho; \mathbf{R}_{nu}) + V^{nl}_{\mathrm{ext}}(\rho; \mathbf{R}_{nu})\,,
\end{equation}
where $V^{loc}_{\mathrm{eff}}$ denotes the local part of the Kohn-Sham effective potential and $V^{nl}_{\mathrm{ext}}$ denotes the non-local part.  The local part includes the Hartree potential $V_{\mathrm{H}}$, the exchange-correlational functional potential $V_{\mathrm{XC}}$, and the local part of the external pseudopotential $V^{loc}_{\mathrm{ext}}$: 
\begin{equation}
     V^{loc}_{\mathrm{eff}}(\rho; \mathbf{R}_{nu}) = V_\mathrm{H}(\rho) + V_{\mathrm{XC}}(\rho) + V_{\mathrm{ext}}^{loc}(\rho; \mathbf{R}_{nu})\,.
\end{equation}

\subsection{Hartree Potential}
The Hartree potential $V_{\mathrm{H}}$ is computed following the tensor-structured approach presented in~\cite{KHOROMSKIJ20095749, Hackbusch_kernel_approx}. 
We recall that the Hartree potential is given by the convolution integral
\begin{equation}
    V_\mathrm{H}(\rho) = \int_{\mathbb{R}^3}\frac{\rho(\mathbf{x'})}{\abs{\mathbf{x}-\mathbf{x'}}}d\mathbf{x'}\,.
    \label{eqnsi:hartree_pot_def}
\end{equation}
We first compute a low-rank Tucker decomposition of the electron density as 
\begin{equation}
    \tilde{\rho}(\mathbf{x}) = \sum_{r_1, r_2, r_3 = 1}^{R_\rho} g_{r_1 r_2 r_3} \varrho^{r_1}_{1}(x_1) \varrho^{r_2}_{2}(x_2) \varrho^{r_3}_{3}(x_3)\,,
    \label{eqnsi:rho_decomp}
\end{equation} 
In the above, $R_\rho$ represents the Tucker rank associated with the low-rank approximation of the electron density. We note that the approximation error decays exponentially with the Tucker rank. In general, the Tucker rank can be chosen to be different along the three Cartesian directions. However, for the sake of simplicity, the ideas are presented here using a uniform rank along the different Cartesian directions.\\
The kernel $\frac{1}{\abs{\mathbf{x} - \mathbf{x'}}}$ in Eq.~\ref{eqnsi:hartree_pot_def} is approximated by a series of Gaussian functions to take advantage of the tensor-structured nature as
\begin{equation}
    \frac{1}{|\mathbf{x}|} \approx \sum_{k=1}^{K} w_k e^{-\alpha_k(x_1^2+x_2^2+x_3^2)}\,,
    \label{eqnsi:kernel_approx}
\end{equation}
where $w_k$ and $\alpha_k$ are coefficients and $K$ is the number of terms used to expand the kernel. We refer to prior works~\cite{Hackbusch_kernel_approx, LowRank_Potential_approx} for the derivation and the algorithm. The pre-computed coefficients can be found on the webpage~\cite{kernel_approx_chart}.
Substituting Eq.~\ref{eqnsi:rho_decomp} and Eq.~\ref{eqnsi:kernel_approx} into Eq.~\ref{eqnsi:hartree_pot_def} , the Hartree potential can be evaluated using a separable form
\begin{equation}
        \Tilde{V}_\mathrm{H}(\mathbf{x})=\sum_{k=1}^K w_k\sum_{r_1, r_2, r_3=1}^{R_\rho} g_{r_1 r_2 r_3} \int \varrho^{r_1}_{1} (x_1') e^{-\alpha_k(x_1-x_1')^2} dx_1' 
        \int \varrho^{r_2}_{2} (x_2') e^{-\alpha_k (x_2-x_2')^2} dx_2' \int \varrho^{r_3}_{3} (x_3') e^{-\alpha_k(x_3-x_3')^2} dx_3'\,.
\end{equation}   

The local part of the effective potential is then computed by summing the Hartree potential, exchange-correlational functional and the local part of the external potential. We next compute the Tucker decomposition of the local part of the effective potential to exploit the tensor structure in the computation of the Hamiltonian matrix elements in the Tucker tensor basis (cf. section~\ref{SI:HamiltonianMatrix}). The Tucker decomposed effective potential is thus represented as

\begin{equation}
    \tilde{V}^{loc}_{\mathrm{eff}} (\mathbf{x}) = \sum_{r_1, r_2, r_3=1}^{R_V} \sigma^V_{r_1r_2r_3} u^{r_1}_{1}(x_1) u^{r_2}_{2}(x_2) u^{r_3}_{3}(x_3)
    \label{eqnsi:decomposed_veff_loc}\,,
\end{equation}
where $R_V$ is the Tucker rank of the local part of the Kohn-Sham effective potential.

\subsection{Non-local projector of pseudopotential}
 In this work, we use the norm-conserving Troullier Martin pseudopotential in Kleinmann-Bylander form. The pseudopotential operator $V_{\mathrm{ext}}$ comprises of a local part $V_{\mathrm{ext}}^{loc}$ and a non-local part $V_{\mathrm{ext}}^{nl}$. The action of the pseudopotential operator on a Kohn-Sham orbital in real-space is given by
\begin{equation}
    V_{\mathrm{ext}}(\mathbf{x};\mathbf{R}_{nu})\Psi_i(\mathbf{x}) = V_{\mathrm{ext}}^{loc}(\mathbf{x};\mathbf{R}_{nu})\Psi_i(\mathbf{x}) + V_{\mathrm{ext}}^{nl}(\mathbf{x};\mathbf{R}_{nu})\Psi_i(\mathbf{x})\,,
    \label{eqnsi:psp_vnl_vloc}
\end{equation}
\begin{equation}
    V_{\mathrm{ext}}^{loc}(\mathbf{x};\mathbf{R}_{nu})\Psi_i(\mathbf{x}) = \sum_{J=1}^{N_a}V_{\mathrm{ext}}^{loc, J}(\mathbf{x}-\mathbf{R}_J)\Psi_i(\mathbf{x})\,
    \label{eqnsi:psp_vloc}
\end{equation}
where $V_{\mathrm{ext}}^{loc, J}(\mathbf{x}-\mathbf{R}_J)$ is the corresponding local potential for the $J$-th atom, and $\mathbf{R}_J$ is the coordinates of the $J$-th atom. The action of the non-local operator in real-space is given by
\begin{equation}
    V_{\mathrm{ext}}^{nl}(\mathbf{x};\mathbf{R}_{nu}) \Psi_i(\mathbf{x}) = \sum_J^{N_a}\sum_{lm}C^J_{lm}\varphi^J_{lm}(\mathbf{x}-\mathbf{R}_J)\Delta V^J_l (\mathbf{x}-\mathbf{R}_J)\,,
    \label{eqn:psp_vnl}
\end{equation}
where
\begin{equation*}
\begin{aligned}
    C^J_{lm}&=\frac{\int \varphi^J_{lm}(\mathbf{x}-\mathbf{R}_J)\Delta V^J_l (\mathbf{x}-\mathbf{R}_J) \Psi_i(\mathbf{x})d\mathbf{x}}{\int \varphi^J_{lm}(\mathbf{x}-\mathbf{R}_J)\Delta V^J_l (\mathbf{x}-\mathbf{R}_J) \varphi^J_{lm}(\mathbf{x}-\mathbf{R}_J) d\mathbf{x}} \\
    &= \frac{1}{\nu^J_{lm}}\int \varphi^J_{lm}(\mathbf{x}-\mathbf{R}_J)\Delta V^J_l (\mathbf{x}-\mathbf{R}_J) \varphi^J_{lm}(\mathbf{x}-\mathbf{R}_J) d\mathbf{x}\,.
\end{aligned}
\end{equation*}
Therein, $\Delta V^J_l (\mathbf{x}-\mathbf{R}_J) =  V^J_l (\mathbf{x}-\mathbf{R}_J) - V_{\mathrm{ext}}^{loc, J}(\mathbf{x}-\mathbf{R}_J)$ is the difference between the pseudopotential component of the $J$-th atom corresponding to the $l$ azimuthal quantum number and the local part of the pseudopotential, and $\varphi^J_{lm}(\mathbf{x})$ is the single atom pseudo-wavefunction of the $J$-th atom corresponding to the $l$ azimuthal quantum number and $m$ magnetic quantum number, and $\nu^J_{lm} \vcentcolon= {\int \varphi^J_{lm}(\mathbf{x}-\mathbf{R}_J) \Delta V^J_l (\mathbf{x}-\mathbf{R}_J) \varphi^J_{lm}(\mathbf{x}-\mathbf{R}_J) d\mathbf{x}}$. \\  
In order to efficiently compute the action of non-local projector on the Kohn-Sham wavefunction,  we introduce an intermediate term $\Lambda^J_{lm}(\mathbf{x}) = \varphi^J_{lm}(\mathbf{x}-\mathbf{R}_J) \Delta V^J_l (\mathbf{x}-\mathbf{R}_J)$ and compute its Tucker decomposition denoted as 
\begin{equation}
    \tilde{\Lambda}^J_{lm}(\mathbf{x}) = \sum_{r_1, r_2, r_3 = 1}^{R_V^{nl}}\sigma^{\Lambda^J_{lm}}_{r_1r_2r_3} \phi^{\Lambda^J_{lm}, r_1}_{1}(x_1) \phi^{\Lambda^J_{lm}, r_2}_{2}(x_2) \phi^{\Lambda^J_{lm}\,, r_3}_{3}(x_3),
    \label{eqnsi:lambda}
\end{equation}
where $R_V^{nl}$ is the associated Tucker rank, chosen to be the largest among the $J$ atoms and its corresponding quantum numbers $l$ and $m$.

The non-local part of the external potential computed with the Tucker decomposed quantities is denoted as $\tilde{V}^{nl}_{\mathrm{ext}}(\mathbf{x})$, and given

\begin{equation}
    \tilde{V}^{nl}_{\mathrm{ext}}(\mathbf{x}) \Psi_i(\mathbf{x}) =  \sum_J^{N_a}\sum_{lm} \tilde{C}^J_{lm} \tilde{\Lambda}^J_{lm}(\mathbf{x})\,,
    \label{eqnsi:decomposed_vext_nonloc}
\end{equation}
where 
\begin{equation*}
    \tilde{C}^J_{lm} = \frac{1}{\nu^J_{lm}}\int \tilde{\Lambda}^J_{lm}(\mathbf{x}) \Psi_i(\mathbf{x})d\mathbf{x} \,.
\end{equation*}

\section{Minimization problem for separable Hamiltonian quantities computation}
In the Lagrangian of the minimization problem  (Eq.~2 in main text), we use the Tucker decomposed effective potential for efficient evaluation of the ensuing integrals. The Lagrangian, accounting for the  normality constraint with the Lagrange multiplier $\lambda$, is given by

\begin{equation}
\begin{aligned}
    L(\Psi')
    =& \int_{\Omega} \left[ \sum_{\ell=1}^{3} \frac{1}{2} \abs{\frac{d\psi_{\ell}(x_\ell)}{dx_{\ell}}}^2 \prod_ {m \neq \ell}^{3} \psi_{m}^2(x_m) + \left(\tilde{V}_{\mathrm{eff}}^{loc} (\mathrm{x}) + \lambda \right) \prod_{\ell = 1}^3\psi_{\ell}^2(x_\ell) + \prod_{\ell=1}^3 \psi_{\ell}(x_\ell)  \tilde{V}_{\mathrm{ext}}^{nl} (\mathbf{x}) \prod_{\ell = 1}^3\psi_{\ell}(x_\ell) \right] d\mathbf{x} \,.
\end{aligned}
    \label{eqnsi:1dlagrangian}
\end{equation}

Taking the variation of the Lagrangian Eq.~\ref{eqnsi:1dlagrangian} with respect to $\psi_{\ell}$, we arrive at the simultaneous 1-D eigenvalue problems 
\begin{equation}
    \left(-\frac{1}{2} \frac{d^2}{dx_k^2} + v^{loc}_k(x_k; \psi_{l \neq k}) + v^{nl}_k(x_k; \psi_{l \neq k})\right)\psi_k(x_k) = -(\lambda + a_k)\psi_k(x_k) \quad\quad k=1,2,3 \,.
    \label{eqnsi:1dks}
\end{equation}
We note that $\mathcal{H}_k=-\frac{1}{2} \frac{d^2}{dx_k^2} + v^{loc}_k(x_k; \psi_{l \neq k}) + v^{nl}_k(x_k; \psi_{l \neq k})$ and $\alpha_k = -(\lambda + a_k)$ in Eq. 3 of the main text. Here, we write out the 1-D quantities in Eq.~\ref{eqnsi:1dks} and refer to Motamarri et al.\cite{Motamarri2016a} for details:
\begin{equation}
    v^{loc}_k(x_k; \psi_{l \neq k}) =  \frac{1}{m_k}\int \tilde{V}^{loc}_{\mathrm{eff}} \prod_{p=1, p\neq k}^3 \psi_p^2(x_p) d \mathbf{\hat{x}_k} \,,
\end{equation}
where $d\mathbf{\hat{x}_k} = \prod_{p=1, p \neq k}^3 d\mathbf{x_p}$, $m_k = \int \prod_{p=1, p\neq k}^3 \psi_p^2 d \mathbf{\hat{x}_k}$,

\begin{equation}
    v^{nl}_k(x_k; \psi_{l \neq k}) \psi_k(x_k) = \frac{1}{m_k}\sum_J^{N_a}\sum_{lm} \tilde{C}^J_{lm} \int \tilde{\Lambda}^J_{lm}(\mathbf{x}) \prod_{p=1, p\neq k}^3 \psi_p(x_p) d \mathbf{\hat{x}_k}\,,
\end{equation}

\begin{equation}
    a_k = \frac{1}{2m_k}\int \sum_{\mathclap{\substack{p, q=1\\  p, q \neq k}}}^3 \hspace{0.15cm} \abs{\frac{d\psi_{p}(x_p)}{dx_p}}^2\psi^2_q(x_q) d\mathbf{\hat{x}_k}\,.
\end{equation}
The minimization problem can thus be written into a set of simultaneous 1-D eigenvalue problem, where each eigenvalue problem is parametrized by the solution of the other two directions. The simultaneous eigenvalue problem can be solved using a self-consistent iteration procedure.

\section{SOC Algorithm} The splitting orthogonality constraint algorithm (SOC) \cite{SOC, Compressed_mode_osher} can be used for finding a set of localized functions which closely approximate the eigenspace, yet preserving the orthogonality of the localized functions. In this work, the SOC algorithm is used to construct localized functions that closely approximate the eigensubspace of the separable approximation of the Kohn-Sham Hamiltonian. Since the 1-D separable Hamiltonian is computed using a finite-element discretization, which in turn results in a generalized eigenvalue problem, the eigenfunctions are $\mathbf{M}$-orthogonal ($\mathbf{M}$ denoting the overlap matrix). We hereby summarize the SOC algorithm for the $\mathbf{M}$-orthogonality constraint variance.

Consider the generalized 1-D eigenvalue problem along direction $k$ given by $\mathbf{H}_k \mathbf{\Psi}_k =  \mathbf{M}_k \mathbf{\Psi}_k \mathbf{\Lambda}_k$, where $\mathbf{\Psi}_k\in \mathbb{R}^{n \times N_k}$ with $n$ denoting the dimension of the basis and $N_k$ denoting the number of lowest eigenstates of interest, and $\mathbf{M}_k$ is the overlap matrix of the k-th dimension. Using Cholesky factorization applied on $\mathbf{M}_k$ so that $\mathbf{M}_k=\mathbf{L}^T\mathbf{L}$, the orthogonality constraint is given by $(\mathbf{L}\mathbf{\Psi}_k)^T (\mathbf{L}\mathbf{\Psi}_k) = \mathbf{I}$. Thus, the original orthogonality constraint for $\mathbf{\Psi}_k$ is thus replaced by $\mathbf{L}\mathbf{\Psi}_k$. The SOC algorithm for a generalized eigenvalue problem is given by
\begin{algorithm}
\SetAlgoLined
\KwIn{$\mathbf{H}_k$, $\mathbf{M}_k$, $\mu$, $\eta$, $\kappa$, tol}
\KwOut{$\mathbf{\Psi}^L_k$}
\textbf{Chelosky factorization:} $\mathbf{M}_k=\mathbf{L}^T\mathbf{L}$\\
\textbf{Initialize:} $\mathbf{P^0}=\mathbf{L} \mathbf{\Psi}_k$, $\mathbf{Q^0}=\mathbf{\Psi}_k$, $\mathbf{b^0}=\mathbf{B^0}=\mathbf{0}$\\
\While{$e > \textrm{tol}$}{
	1. $\mathbf{\Psi}^i_{k} = \arg \min_{\mathbf{\Psi}'_k} \textrm{Tr}(\mathbf{\Psi}'{}^T_k \mathbf{H}_k \mathbf{\Psi}'_k) + \frac{\eta}{2}\|\mathbf{\Psi}'_k-\mathbf{Q}^{i-1}+\mathbf{b}^{i-1}\|^2_F + \frac{\kappa}{2} \|\mathbf{L} \mathbf{\Psi}'_k-\mathbf{P}^{i-1}+\mathbf{B}^{i-1}\|^2_F$ \\
	2. $\mathbf{Q}^i = \arg \min_\mathbf{Q} \frac{1}{\mu}\abs{\mathbf{Q}} + \frac{\eta}{2}\|\mathbf{\Psi}{}^i_k-\mathbf{Q}+\mathbf{b}^{i-1}\|^2_F$ \\
	3. $\mathbf{P}^i = \arg \min_\mathbf{P} \|\mathbf{L}\mathbf{\Psi}{}^i_k-\mathbf{P}+\mathbf{B}^{i-1}\|^2_F \textrm{ s.t. } \mathbf{P}^T\mathbf{P}=\mathbf{I}$ \\
	4. $\mathbf{b}^i = \mathbf{b}^{i-1} + \mathbf{\Psi}{}^i_k - \mathbf{Q}^i$ \\
	5. $\mathbf{B}^i = \mathbf{B}^{i-1} + \mathbf{L}\mathbf{\Psi}{}^i_k - \mathbf{P}^i$ \\
	6. $E^i = \frac{1}{\mu} \abs{\mathbf{\Psi}^i_k} + \mathrm{Tr}({\mathbf{\Psi}^i_k}^T \mathbf{H}_k \mathbf{\Psi}^i_k)$ \\
	7. $e = \abs{\frac{E^i - E^{i-1}}{E^{i}}}$  \\
	8. \textbf{if} $e < \mathrm{tol}$; \textbf{then} $\mathbf{\Psi}^L_{k} = \mathbf{\Psi}^i_k$
	}
\caption{SOC}
\label{alg:SOC}
\end{algorithm}
 
In the Algorithm \ref{alg:SOC}, $\eta$ and $\kappa$ are the penalty factors for each constraint, tol is the stopping criteria for the error measure $e$. $\mathbf{\Psi}^L_{k}$ is the computed localized 1-D functions. The solution for the three sub-problems 1-3 are
\begin{equation}
	\begin{aligned}
		&2(\mathbf{H}_k+\eta+\kappa \mathbf{L}) \mathbf{\Psi}{}^i_k=\kappa(\mathbf{P}^{i-1}-\mathbf{B}^{i-1})+\eta(\mathbf{Q}^{i-1}-\mathbf{b}^{i-1})\\
		&\mathbf{Q}^i = \textrm{sign}(\mathbf{\Psi}{}^i_k+\mathbf{b}^{i-1})\max{\left(0, \abs{\mathbf{\Psi}{}^i_k+\mathbf{b}^{i-1}}-\frac{1}{\eta\mu}\right)}\\
		&\mathbf{P}^i = (\mathbf{L}\mathbf{\Psi}{}^i_k + \mathbf{B}^{i-1})\mathbf{U}\mathbf{S}^{-\frac{1}{2}}\mathbf{V}^T
	\end{aligned},
	\label{SOC_solution}
\end{equation}
where $\mathbf{U}$, $\mathbf{V}$, $\mathbf{S}$ are the left singular vectors, the right singular vectors, and the singular values for $(\mathbf{L} \mathbf{\Psi}{}^i_k + \mathbf{B}^{i-1})^T(\mathbf{L} \mathbf{\Psi}{}^i_k + \mathbf{B}^{i-1})$, respectively.

\section{Computation of projected Hamiltonian}\label{SI:HamiltonianMatrix}
Using the effective potential decomposed into the Tucker tensor format, the computation of the projected Hamiltonian is elaborated here for clarity.

Substituting the decomposed quantities Eq.~\ref{eqnsi:decomposed_veff_loc}, Eq.~\ref{eqnsi:decomposed_vext_nonloc} in Eq. 6 in the main text, the entries of the Hamiltonian matrix in the localized Tucker tensor basis are given by 
\begin{equation}
\begin{aligned}
	\tilde{H}^L_{I, J} =& \Bra{T^L_{I}}  -\frac{1}{2}\nabla^2+ \tilde{V}^{loc}_{\mathrm{eff}} +  \tilde{V}^{nl}_{\mathrm{ext}} \Ket{T^L_{J}} \\
	=& \frac{1}{2}\int \nabla T^L_I \cdot \nabla T^L_J d\mathbf{x} + \int T^L_I \tilde{V}^{loc}_{\mathrm{eff}} T^L_J d\mathbf{x} + \int T^L_I \tilde{V}^{nl}_{\mathrm{ext}} T^L_J d\mathbf{x} \,.
\end{aligned}
\label{eqnsi:projected hamiltonian}
\end{equation}
Recall that $T^L_I = \psi_{1, i_1}^L(x_1) \psi_{2, i_2}^L(x_2) \psi_{3, i_3}^L(x_3)$ has a tensor structure, and thus each term in Eq.~\ref{eqnsi:projected hamiltonian} can be computed using 1-D integrals as follows:

\begin{equation}
\begin{aligned}
    \int \nabla T^L_I \cdot \nabla T^L_J d\mathbf{x} =& \int \nabla (\psi_{1, i_1}^L \psi_{2, i_2}^L \psi_{3, i_3}^L) \cdot \nabla ( \psi_{1, j_1}^L \psi_{2, j_2}^L \psi_{3, j_3}^L) \,d\mathbf{x} \\
    =& \,\,G^{dx_1}_{i_1j_1}G^{x_2}_{i_2j_2}G^{x_3}_{i_3j_3} + G^{x_1}_{i_1j_1}G^{dx_2}_{i_2j_2}G^{x_3}_{i_3j_3} + G^{x_1}_{i_1j_1}G^{x_2}_{i_2j_2}G^{dx_3}_{i_3j_3}\,,
\end{aligned}
\end{equation}
where 
\begin{equation*}
    G^{dx_d}_{i_d j_d} = \int \frac{d \psi_{d, i_d}^L(x_d)}{dx_d} \frac{d \psi_{d, j_d}^L(x_d)}{dx_d}dx_d
\end{equation*}
and
\begin{equation*}
    G^{x_d}_{i_d j_d} = \int \psi_{d, i_d}^L(x_d)\psi_{d, j_d}^L(x_d) dx_d.
\end{equation*}

\begin{equation}
\begin{aligned}
    \int T^L_I \tilde{V}^{loc}_{\mathrm{eff}} T^L_J d\mathbf{x} =& \int \psi_{1, i_1}^L(x_1) \psi_{2, i_2}^L(x_2) \psi_{3, i_3}^L(x_3) \left(\sum_{r_1, r_2, r_3 = 1}^{R_V} \sigma^V_{r_1r_2r_3} u^{r_1}_{1}(x_1) u^{r_2}_{2}(x_2) u^{r_3}_{3}(x_3)\right) \\
    &\psi_{1, j_1}^L(x_1) \psi_{2, j_2}^L(x_2) \psi_{3, j_3}^L(x_3) d\mathbf{x}\\ 
    =& \sum_{r_1, r_2, r_3 = 1}^{R_V} \sigma^V_{r_1r_2r_3} \int\psi_{1, i_1}^L(x_1) u^{r_1}_{1}(x_1) \psi_{1, i_1}^L(x_1) dx_1 \int\psi_{2, i_2}^L(x_2) u^{r_2}_{2}(x_2) \psi_{2, i_2}^L(x_2) dx_2 \\
    &\int\psi_{3, i_3}^L(x_3) u^{r_3}_{3}(x_3) \psi_{3, i_3}^L(x_3) dx_3\,. 
\end{aligned}
\label{eqnsi:eff pot project}
\end{equation}

For the non-local part of the effective potential, we first consider the expression

\begin{equation}
\begin{aligned}
    &\int T^L_I V^{nl}_{\mathrm{ext}}(\mathbf{x}) T^L_J d\mathbf{x}  \\
    =& \sum_J^{N_a}\sum_{lm} \frac{1}{\nu^J_{lm}}\int \psi_{1, i_1}^L\psi_{2, i_2}^L \psi_{3, i_3}^L  \Lambda^J_{lm}(\mathbf{x}) d\mathbf{x} \int \Lambda^J_{lm}(\mathbf{x}) \psi_{1, j_1}^L\psi_{2, j_2}^L \psi_{3, j_3}^L d\mathbf{x} \,.
\end{aligned}
\label{eqnsi:potnl_non_decomp}
\end{equation}
We note that the right-hand side of equation Eq.~\ref{eqnsi:potnl_non_decomp} is a matrix operation $Q^T Q$, where

\begin{equation}
    Q_{Jlm, i_1i_2i_3} = \int \psi_{1, i_1}^L(x_1)\psi_{2, i_2}^L(x_2) \psi_{3, i_3}^L(x_3)  \Lambda^J_{lm}(\mathbf{x}) d\mathbf{x}.
\end{equation}
Using the low-rank Tucker decomposition of $\Lambda^J_{lm}(\mathbf{x})$ in Eq.~\ref{eqnsi:lambda},
\begin{equation*}
    \tilde{\Lambda}^J_{lm}(\mathbf{x}) = \sum_{r_1, r_2, r_3 = 1}^{R_V^{nl}}\sigma^{\Lambda^J_{lm}}_{r_1r_2r_3} \phi^{\Lambda^J_{lm}, r_1}_{1}(x_1) \phi^{\Lambda^J_{lm}, r_2}_{2}(x_2) \phi^{\Lambda^J_{lm}, r_3}_{3}(x_3)\,,
\end{equation*}

$Q_{Jlm, i_1i_2i_3}$ can be computed  in a tensor decomposed form~ using $\tilde{\Lambda}^J_{lm}$ as

\begin{equation*}
     \tilde{Q}_{Jlm, i_1i_2i_3} = \sum_{r_1, r_2, r_3}^{R_V^{nl}}\sigma^{\Lambda^J_{lm}}_{r_1r_2r_3} \int \phi^{\Lambda^J_{lm}, r_1}_{1}(x_1) \psi_{1, i_1}^L(x_1) dx_1 \int \phi^{\Lambda^J_{lm}, r_2}_{2}(x_2) \psi_{2, i_2}^L(x_2) dx_2 \int \phi^{\Lambda^J_{lm}, r_3}_{3}(x_3) \psi_{3, i_3}^L(x_3) dx_3\,.
\end{equation*}

Thus, the non-local part of the projected Hamiltonian in the 3-D localized tensor-structured format with $\tilde{V}^{nl}_{\mathrm{ext}}$ is given by 
\begin{equation}
    \int T^L_I \tilde{V}^{nl}_{\mathrm{ext}}(\mathbf{x}) T^L_J d\mathbf{x} = \frac{1}{\nu^J_{lm}}\tilde{Q}^T\tilde{Q} \,.
\end{equation}

\section{The localized 1-D functions for $  \mathrm{Al}_{147}$}
We computed the additive separable approximation of the Kohn-Sham Hamiltonian $\mathcal{H}_k$ for the $\mathrm{Al}_{147}$ nano-particle with icosahedral symmetry, and, using this, computed the lowest 70 eigenstates of $\mathcal{H}_k$. We subsequently use the $L_1$ localization approach to compute the localized functions that are a close approximation to the eigensubspace. Figure ~\ref{Fig:70 states} shows the lowest 70 eigenfunctions of $\mathcal{H}_1$ (top) and the corresponding 1-D localized functions (bottom). It is evident that while the eigenfunctions are global, the $L_1$ localized functions are fairly localized in space. As noted in the main text, the locality is the key to constructing a sparse Kohn-Sham Hamiltonian, and the resulting computational efficiency.

\section{Adaptive vs. fixed Tucker tensor basis}
As the electron density evolves during the self-consistent field (SCF) iteration, the Kohn-Sham Hamiltonian changes, and thus the additive separable approximation to Kohn-Sham Hamiltonian also changes with SCF iteration. Here, we study the difference in the approximation properties of the adaptive Tucker tensor basis---one which is constructed as the eiegnbasis of the additive separable approximation to the Kohn-Sham Hamiltonian in every SCF iteration---with those of a fixed Tucker tensor basis where the basis is constructed only in the first SCF iteration (using an input electron density being a superposition of atomic densities) and held fixed during the course of the SCF iteration. To this end, we consider two benchmark systems: (i) $\mathrm{Al_{147}}$; (ii) $\mathrm{Si_{220}H_{144}}$. Table~\ref{tab:rank_vs_basis_fixed} compares the approximation properties of the adaptive and the fixed Tucker basis. Notably, the difference in the ground-state energies computed using the adaptive and fixed Tucker tensor basis for any given rank is significantly smaller than the basis discretization error corresponding to the rank. To elaborate, the difference in ground-state energies between the adaptive and fixed Tucker tensor basis for rank 70 for $\mathrm{Al_{147}}$ is $<1$ meV/atom, whereas the basis discretization error corresponding to this rank is $\sim 8.5$ meV/atom (Reference energy is -56.6274 eV/atom). Similarly, for $\mathrm{Si_{220}H_{144}}$, the difference between the adaptive and fixed basis is again $<1$ meV/atom for rank 80, whereas the basis discretization error is $\sim 7.5$ meV/atom (Reference energy is -71.393 eV/atom). These results suggest that it suffices to use a fixed Tucker tensor basis. 

\section{Truncation tolerance in computation of projected Hamiltonian $\mathbf{H}^{L}$}
As discussed in the main text, a truncation tolerance is used to zero the elements below a threshold in the projected Hamiltonian (i.e., the Kohn-Sham Hamiltonian projected onto $\mathbb{T}^{L}$). We present data to quantify the error introduced by this truncation in the computed ground-state energy. To this end, we again consider the two benchmark problems: (a) $\mathrm{Al_{147}}$ and (b) $\mathrm{Si_{220}H_{144}}$. Table~\ref{tab:truncation_tolerance} provides the ground-state energies for various values of the truncation tolerance, computed using the 3-D localized Tucker tensor basis with Tucker rank 70 for $\mathrm{Al_{147}}$ and rank 80 for $\mathrm{Si_{220}H_{144}}$. Firstly, we note that there is a systematic decrease in the error with decreasing truncation tolerance. Notably, a truncation tolerance of 1e-4 results in a $\sim 1$ meV/atom error, which is significantly lower than the basis discretization error and the desired accuracy, but provides excellent sparsity in $\mathbf{H}^{L}$. Thus, a truncation tolerance of 1e-4 has been adopted in benchmark studies on aluminum nano-particles and silicon quantum dots.

\section{Breakdown of computational times for solving the Kohn-Sham equations}
Table~\ref{tab:time breakdown} shows the breakdown of computational times for the various steps involved in solving the Kohn-Sham equations in the localized Tucker tensor basis using the Chebyshev filtering based subspace iteration (ChFSI)~\cite{ChFSIoriginal}. This breakdown is provided for all the benchmark problems involving aluminum nano-particles and silicon quantum dots. As presented in the main text, ChFSI comprises of: (i) Chebyshev filtering (denoted as ChF) to construct a subspace that is a close approximation to the desired eigenspace of the occupied states; (ii) orthogonalization of Chebyshev filtered vectors (denoted as Orth); (iii) projecting the Kohn-Sham Hamiltonian onto the Chebyshev filtered orthogonalized subspace (denoted as Sub proj), and (4) diagonalization followed by other calculations to proceed to the next iteration in the SCF (denoted as Others). As is evident from the data in Table~\ref{tab:time breakdown}, Chebyshev filtering is the dominant cost for all the systems, even for those with 20,000-25,000 electrons. The subquadratic scaling with system size that is observed in this study is a consequence of the slow increase in the Tucker rank with system size that results in a sublinear scaling of the number of Tucker basis functions with system size (cf. Table 1 in main text). We also note that the sparsity of $\mathbf{H}^{L}$ realized using a truncation tolerance of 1e-4 either improves or is maintained with increasing system size. 

\clearpage

\begin{table}[!htb]
\centering
\begin{tabular}{ccccccccc}
\hline
\multicolumn{1}{|c|}{{}}                                                                                     & \multicolumn{4}{c|}{{\textbf{Tucker}}}                                                                                                                                                                                                                                         & \multicolumn{3}{c|}{{\textbf{QE}}}                                                                                                                                                                                 & \multicolumn{1}{c|}{{}}                                       \\ \cline{2-8}
\multicolumn{1}{|c|}{\multirow{-2}{*}{{\textbf{\begin{tabular}[c]{@{}c@{}}Nano\\ particle\end{tabular}}}}}   & \multicolumn{1}{c|}{{\textbf{rank}}} & \multicolumn{1}{c|}{{\textbf{\begin{tabular}[c]{@{}c@{}}\# basis/\\ atom\end{tabular}}}} & \multicolumn{1}{c|}{{\textbf{E/atom}}} & \multicolumn{1}{c|}{{\textbf{time}}} & \multicolumn{1}{c|}{{\textbf{\begin{tabular}[c]{@{}c@{}}\#basis/\\ atom\end{tabular}}}}  & \multicolumn{1}{c|}{{\textbf{E/atom}}} & \multicolumn{1}{c|}{{\textbf{time}}} & \multicolumn{1}{c|}{\multirow{-2}{*}{{\textbf{\begin{tabular}[c]{@{}c@{}}Reference\\ Energy/atom\end{tabular}}}}} \\ \hline
\multicolumn{1}{|c|}{{$\mathrm{Al}_{13}$}}                                                                   & \multicolumn{1}{c|}{{40}}            & \multicolumn{1}{c|}{{4923}}                                                              & \multicolumn{1}{c|}{{-55.9965}}        & \multicolumn{1}{c|}{{0.00067}}       & \multicolumn{1}{c|}{{12403}}                                                             & \multicolumn{1}{c|}{{-55.9993}}        & \multicolumn{1}{c|}{{0.00022}}       & \multicolumn{1}{c|}{{-56.0034*}}                              \\ \hline
\multicolumn{1}{|c|}{{$\mathrm{Al}_{147}$}}                                                                  & \multicolumn{1}{c|}{{70}}            & \multicolumn{1}{c|}{{2333}}                                                              & \multicolumn{1}{c|}{{-56.6179}}        & \multicolumn{1}{c|}{{0.069}}         & \multicolumn{1}{c|}{{5078}}                                                              & \multicolumn{1}{c|}{{-56.6198}}        & \multicolumn{1}{c|}{{0.028}}         & \multicolumn{1}{c|}{{-56.6274*}}                              \\ \hline
\multicolumn{1}{|c|}{{$\mathrm{Al}_{561}$}}                                                                  & \multicolumn{1}{c|}{{85}}            & \multicolumn{1}{c|}{{1095}}                                                              & \multicolumn{1}{c|}{{-56.8119}}        & \multicolumn{1}{c|}{{0.69}}          & \multicolumn{1}{c|}{{4490}}                                                              & \multicolumn{1}{c|}{{-56.8122}}        & \multicolumn{1}{c|}{{1.24}}          & \multicolumn{1}{c|}{{-56.8191*}}                              \\ \hline
\multicolumn{1}{|c|}{{$\mathrm{Al}_{2057}$}}                                                                 & \multicolumn{1}{c|}{{120}}           & \multicolumn{1}{c|}{{840}}                                                               & \multicolumn{1}{c|}{{-56.9192}}        & \multicolumn{1}{c|}{{7.96}}          & \multicolumn{1}{c|}{{5015 }}                                                              & \multicolumn{1}{c|}{{N/A}}             & \multicolumn{1}{c|}{{66.68}}         & \multicolumn{1}{c|}{{-56.9284$^\dagger$}}             \\ \hline
\multicolumn{1}{|c|}{{$\mathrm{Al}_{6525}$}}                                                                 & \multicolumn{1}{c|}{{150}}           & \multicolumn{1}{c|}{{517}}                                                               & \multicolumn{1}{c|}{{-57.0013}}        & \multicolumn{1}{c|}{{55.08}}         & \multicolumn{1}{c|}{{N/A}}                                                               & \multicolumn{1}{c|}{{N/A}}             & \multicolumn{1}{c|}{{N/A}}           & \multicolumn{1}{c|}{{-57.0090$^\dagger$}}             \\ \hline
\multicolumn{9}{c}{{(a)}}                                                                                                                                                                                                                                                                                                                                                                                                                                                                                                                                                                                                                                                                                                                         \\ \hline
\multicolumn{1}{|c|}{{}}                                                                                     & \multicolumn{4}{c|}{{\textbf{Tucker}}}                                                                                                                                                                                                                                         & \multicolumn{3}{c|}{{\textbf{QE}}}                                                                                                                                                                                 & \multicolumn{1}{c|}{{}}                                       \\ \cline{2-8}
\multicolumn{1}{|c|}{\multirow{-2}{*}{{\textbf{\begin{tabular}[c]{@{}c@{}}Si quantum\\ dots\end{tabular}}}}} & \multicolumn{1}{c|}{{\textbf{rank}}} & \multicolumn{1}{c|}{{\textbf{\begin{tabular}[c]{@{}c@{}}\# basis/\\ atom\end{tabular}}}} & \multicolumn{1}{c|}{{\textbf{E/atom}}} & \multicolumn{1}{c|}{{\textbf{time}}} & \multicolumn{1}{c|}{{\textbf{\begin{tabular}[c]{@{}c@{}}\# basis/\\ atom\end{tabular}}}} & \multicolumn{1}{c|}{{\textbf{E/atom}}} & \multicolumn{1}{c|}{{\textbf{time}}} & \multicolumn{1}{c|}{\multirow{-2}{*}{{\textbf{\begin{tabular}[c]{@{}c@{}}Reference\\ Energy/atom\end{tabular}}}}} \\ \hline
\multicolumn{1}{|c|}{{$\mathrm{Si}_{10}\mathrm{H}_{16}$}}                                                    & \multicolumn{1}{c|}{{45}}            & \multicolumn{1}{c|}{{3505}}                                                              & \multicolumn{1}{c|}{{-51.0271}}        & \multicolumn{1}{c|}{{0.0065}}        & \multicolumn{1}{c|}{{7048}}                                                              & \multicolumn{1}{c|}{{-51.0279}}        & \multicolumn{1}{c|}{{0.00014}}       & \multicolumn{1}{c|}{{-51.0339*}}                              \\ \hline
\multicolumn{1}{|c|}{{$\mathrm{Si}_{220}\mathrm{H}_{144}$}}                                                  & \multicolumn{1}{c|}{{80}}            & \multicolumn{1}{c|}{{1407}}                                                              & \multicolumn{1}{c|}{{-71.3841}}        & \multicolumn{1}{c|}{{0.094}}         & \multicolumn{1}{c|}{{2534}}                                                              & \multicolumn{1}{c|}{{-71.3839}}        & \multicolumn{1}{c|}{{0.096}}         & \multicolumn{1}{c|}{{-71.3930*}}                              \\ \hline
\multicolumn{1}{|c|}{{$\mathrm{Si}_{525}\mathrm{H}_{276}$}}                                                  & \multicolumn{1}{c|}{{90}}            & \multicolumn{1}{c|}{{910}}                                                               & \multicolumn{1}{c|}{{-76.1182}}        & \multicolumn{1}{c|}{{0.96}}          & \multicolumn{1}{c|}{{2251}}                                                              & \multicolumn{1}{c|}{{-76.1194}}        & \multicolumn{1}{c|}{{1.12}}          & \multicolumn{1}{c|}{{-76.1279*}}                              \\ \hline
\multicolumn{1}{|c|}{{$\mathrm{Si}_{1214}\mathrm{H}_{504}$}}                                                 & \multicolumn{1}{c|}{{100}}           & \multicolumn{1}{c|}{{582}}                                                               & \multicolumn{1}{c|}{{-80.8627}}        & \multicolumn{1}{c|}{{3.85}}          & \multicolumn{1}{c|}{{2132}}                                                              & \multicolumn{1}{c|}{{N/A}}             & \multicolumn{1}{c|}{{20.01}}         & \multicolumn{1}{c|}{{-80.8717$^\dagger$}}             \\ \hline
\multicolumn{1}{|c|}{{$\mathrm{Si}_{6047}\mathrm{H}_{1308}$}}                                                & \multicolumn{1}{c|}{{140}}           & \multicolumn{1}{c|}{{373}}                                                               & \multicolumn{1}{c|}{{-91.5659}}        & \multicolumn{1}{c|}{{67.49}}         & \multicolumn{1}{c|}{{N/A}}                                                               & \multicolumn{1}{c|}{{N/A}}             & \multicolumn{1}{c|}{{N/A}}           & \multicolumn{1}{c|}{{-91.5741$^\dagger$}}             \\ \hline
\multicolumn{9}{c}{{(b)}}                                                                                                                                                                                                                                                                                                                                                                                                                                                                                                                                                                                                                                                                                                                        
\end{tabular}
\caption{ Comparison of the computational performance of the tensor-structured approach with Quantum Espresso (QE) for two benchmark systems: (a) Aluminum nano-particles; (b) Silicon quantum dots. All energies are reported in eV, and the computational times are reported in node-hrs per SCF iteration. The plane-wave cut-off employed for QE calculations to target the desired accuracy is 25 Ha for aluminum nano-particles and 20 Ha for silicon quantum dots. Full ground-state calculations were performed using the tensor-structured approach for all systems. In the case of QE, full-ground-state calculations were performed for the systems where the ground-state energies are provided, whereas for $\mathrm{Al_{2057}}$ and $\mathrm{Si_{1214}H_{504}}$ only a few SCF iterations were performed to compute the stable SCF time due to significantly increased computational cost. $\mathrm{Al_{6525}}$ and $\mathrm{Si_{6047}H_{1308}}$ systems were beyond reach using QE. The reference energies are computed using QE (*) with higher plane-wave cut-off for smaller systems---55 Ha for aluminum nano-particles and 50 Ha for silicon quantum dots. The reference energies for the larger systems are obtained using DFT-FE ($\dagger$).
}
\label{tab:full_data}
\end{table}

\clearpage
\clearpage

\renewcommand{\thefigure}{SI\arabic{figure}}
\setcounter{figure}{0}
\begin{figure}[!htb]
\centering
\includegraphics[width=\textwidth]{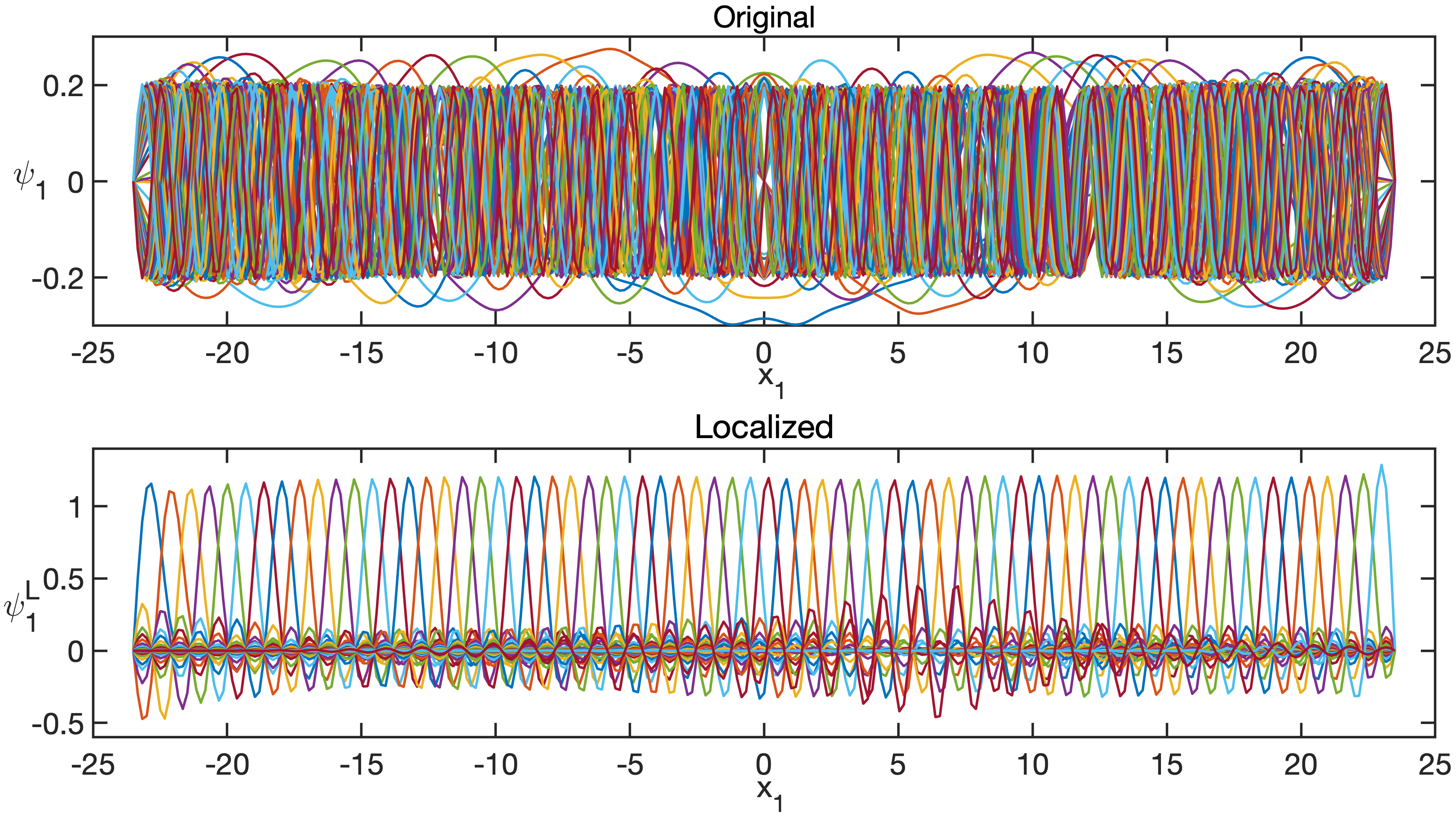}
\caption{1-D functions in $x_1$ direction constructed from the additive separable approximation of the Kohn-Sham Hamiltonian  for $\mathrm{Al}_{147}$ nano-partice. Top: Lowest 70 eigenfunctions of $\mathcal{H}_1$. Bottom: The corresponding $L_1$ localized 1-D functions.}
\label{Fig:70 states}
\end{figure}

\renewcommand{\thetable}{SI\arabic{table}}
\setcounter{table}{0}
\begin{table}[!htb]
\centering
\begin{tabular}{ccccccccc}
\cline{1-4} \cline{6-9}
\multicolumn{1}{|c|}{\textbf{Rank}} & \multicolumn{1}{c|}{\textbf{\begin{tabular}[c]{@{}c@{}}Adaptive\\ basis\end{tabular}}} & \multicolumn{1}{c|}{\textbf{\begin{tabular}[c]{@{}c@{}}Fixed\\ basis\end{tabular}}} & \multicolumn{1}{c|}{\textbf{\begin{tabular}[c]{@{}c@{}}Difference\\ (meV/atom)\end{tabular}}} & \multicolumn{1}{c|}{} & \multicolumn{1}{|c|}{\textbf{Rank}} & \multicolumn{1}{c|}{\textbf{\begin{tabular}[c]{@{}c@{}}Adaptive\\ basis\end{tabular}}} & \multicolumn{1}{c|}{\textbf{\begin{tabular}[c]{@{}c@{}}Fixed\\ basis\end{tabular}}} & \multicolumn{1}{c|}{\textbf{\begin{tabular}[c]{@{}c@{}}Difference\\ (meV/atom)\end{tabular}}} \\ \cline{1-4} \cline{6-9} 
\multicolumn{1}{|c|}{50}            & \multicolumn{1}{c|}{-56.58066}         & \multicolumn{1}{c|}{-56.57687}            & \multicolumn{1}{c|}{3.79}              & \multicolumn{1}{c|}{} & \multicolumn{1}{c|}{50}            & \multicolumn{1}{c|}{-71.05245}         & \multicolumn{1}{c|}{-71.04783}            & \multicolumn{1}{c|}{4.62}              \\ \cline{1-4} \cline{6-9} 
\multicolumn{1}{|c|}{60}            & \multicolumn{1}{c|}{-56.60263}         & \multicolumn{1}{c|}{-56.60099}            & \multicolumn{1}{c|}{1.64}              & \multicolumn{1}{c|}{} & \multicolumn{1}{c|}{60}            & \multicolumn{1}{c|}{-71.27783}         & \multicolumn{1}{c|}{-71.27434}            & \multicolumn{1}{c|}{3.49}              \\ \cline{1-4} \cline{6-9} 
\multicolumn{1}{|c|}{65}            & \multicolumn{1}{c|}{-56.61170}          & \multicolumn{1}{c|}{-56.61028}            & \multicolumn{1}{c|}{1.42}              & \multicolumn{1}{c|}{} & \multicolumn{1}{c|}{70}            & \multicolumn{1}{c|}{-71.36808}         & \multicolumn{1}{c|}{-71.36661}            & \multicolumn{1}{c|}{1.47}              \\ \cline{1-4} \cline{6-9} 
\multicolumn{1}{|c|}{70}            & \multicolumn{1}{c|}{-56.61964}         & \multicolumn{1}{c|}{-56.61893}            & \multicolumn{1}{c|}{0.71}              & \multicolumn{1}{c|}{} & \multicolumn{1}{c|}{80}            & \multicolumn{1}{c|}{-71.38623}         & \multicolumn{1}{c|}{-71.38539}            & \multicolumn{1}{c|}{0.84}              \\ \cline{1-4} \cline{6-9} 
\multicolumn{1}{|c|}{75}            & \multicolumn{1}{c|}{-56.62425}         & \multicolumn{1}{c|}{-56.62381}            & \multicolumn{1}{c|}{0.44}              & \multicolumn{1}{c|}{} & \multicolumn{1}{c|}{90}            & \multicolumn{1}{c|}{-71.38866}         & \multicolumn{1}{c|}{-71.38821}            & \multicolumn{1}{c|}{0.45}              \\ \cline{1-4} \cline{6-9} 
\multicolumn{1}{|c|}{80}            & \multicolumn{1}{c|}{-56.62637}         & \multicolumn{1}{c|}{-56.62608}            & \multicolumn{1}{c|}{0.29}              & \multicolumn{1}{c|}{} & \multicolumn{1}{c|}{100}           & \multicolumn{1}{c|}{-71.39145}         & \multicolumn{1}{c|}{-71.39101}            & \multicolumn{1}{c|}{0.44}              \\ \cline{1-4} \cline{6-9} 
                                    &                                        &                                           &                                          & \multicolumn{1}{c|}{} & \multicolumn{1}{c|}{105}           & \multicolumn{1}{c|}{-71.39201}         & \multicolumn{1}{c|}{-71.39181}            & \multicolumn{1}{c|}{0.20}              \\ \cline{6-9} 
\multicolumn{4}{c}{(a)}                                                                                                                                             &                       & \multicolumn{4}{c}{(b)}                                                                                                                                           
\end{tabular}
\caption{ Comparison of the approximation properties of adaptive and fixed Tucker tensor basis. The per atom ground-state energies are reported in eV for: (a) $\mathrm{Al_{147}}$; (b) $\mathrm{Si_{220}H_{144}}$.}
\label{tab:rank_vs_basis_fixed}
\end{table}

\begin{table}[!htb]
\centering
\begin{tabular}{cccc}
\hline
\multicolumn{1}{|c|}{\textbf{Truncation   tolerance}} & \multicolumn{1}{c|}{\textbf{Matrix sparsity}} & \multicolumn{1}{c|}{\textbf{\begin{tabular}[c]{@{}c@{}}E/atom\\ (eV/atom)\end{tabular}}} & \multicolumn{1}{c|}{\textbf{\begin{tabular}[c]{@{}c@{}}Difference with\\ no truncation (meV/atom)\end{tabular}}} \\ \hline
\multicolumn{1}{|c|}{no truncation}                   & \multicolumn{1}{c|}{N/A}                      & \multicolumn{1}{c|}{-56.61893}          & \multicolumn{1}{c|}{0}                                                                                           \\ \hline
\multicolumn{1}{|c|}{1e-16}                        & \multicolumn{1}{c|}{0.294987723}              & \multicolumn{1}{c|}{-56.61886}          & \multicolumn{1}{c|}{0.07}                                                                                    \\ \hline
\multicolumn{1}{|c|}{1e-12}                        & \multicolumn{1}{c|}{0.652828618}              & \multicolumn{1}{c|}{-56.61859}          & \multicolumn{1}{c|}{0.34}                                                                                    \\ \hline
\multicolumn{1}{|c|}{1e-09}                        & \multicolumn{1}{c|}{0.875758917}              & \multicolumn{1}{c|}{-56.61833}          & \multicolumn{1}{c|}{0.60}                                                                                    \\ \hline
\multicolumn{1}{|c|}{1e-06}                        & \multicolumn{1}{c|}{0.983255385}              & \multicolumn{1}{c|}{-56.61806}          & \multicolumn{1}{c|}{0.87}                                                                                    \\ \hline
\multicolumn{1}{|c|}{1e-04}                        & \multicolumn{1}{c|}{0.996135765}              & \multicolumn{1}{c|}{-56.61791}          & \multicolumn{1}{c|}{1.02}                                                                                    \\ \hline
\multicolumn{1}{|c|}{1e-03}                        & \multicolumn{1}{c|}{0.999547258}              & \multicolumn{1}{c|}{-56.61543}          & \multicolumn{1}{c|}{3.50}                                                                                    \\ \hline
\multicolumn{4}{c}{(a)}                                                                                                                                                                                                                                               \\ \hline
\multicolumn{1}{|c|}{\textbf{Truncation tolerance}}   & \multicolumn{1}{c|}{\textbf{Matrix sparsity}} & \multicolumn{1}{c|}{\textbf{\begin{tabular}[c]{@{}c@{}}E/atom\\ (eV/atom)\end{tabular}}}  & \multicolumn{1}{c|}{\textbf{\begin{tabular}[c]{@{}c@{}}Difference with\\ no truncation (meV/atom)\end{tabular}}} \\ \hline
\multicolumn{1}{|c|}{no truncation}                   & \multicolumn{1}{c|}{N/A}                      & \multicolumn{1}{c|}{-71.38539}          & \multicolumn{1}{c|}{0}                                                                                           \\ \hline
\multicolumn{1}{|c|}{1e-16}                        & \multicolumn{1}{c|}{0.375215542}              & \multicolumn{1}{c|}{-71.38521}          & \multicolumn{1}{c|}{0.18}                                                                                     \\ \hline
\multicolumn{1}{|c|}{1e-12}                        & \multicolumn{1}{c|}{0.743421981}              & \multicolumn{1}{c|}{-71.38493}          & \multicolumn{1}{c|}{0.46}                                                                                     \\ \hline
\multicolumn{1}{|c|}{1e-09}                        & \multicolumn{1}{c|}{0.927431159}              & \multicolumn{1}{c|}{-71.38472}          & \multicolumn{1}{c|}{0.67}                                                                                     \\ \hline
\multicolumn{1}{|c|}{1e-06}                        & \multicolumn{1}{c|}{0.992731562}              & \multicolumn{1}{c|}{-71.38462}          & \multicolumn{1}{c|}{0.77}                                                                                     \\ \hline
\multicolumn{1}{|c|}{1e-04}                        & \multicolumn{1}{c|}{0.999817424}              & \multicolumn{1}{c|}{-71.38409}          & \multicolumn{1}{c|}{1.30}                                                                                     \\ \hline
\multicolumn{1}{|c|}{1e-03}                        & \multicolumn{1}{c|}{0.999932175}              & \multicolumn{1}{c|}{-71.38270}          & \multicolumn{1}{c|}{2.69}                                                                                     \\ \hline
\multicolumn{4}{c}{(b)}                                                                                                                                                                                                                                              
\end{tabular}
\caption{ The effect of truncation tolerance on ground-state energy and sparsity of $\mathbf{H}^{L}$ for: (a) $\mathrm{Al_{147}}$; (b) $\mathrm{Si_{220}H_{144}}$.}
\label{tab:truncation_tolerance}
\end{table}

\begin{table}[!htb]
\centering
\begin{tabular}{ccccccccc}
\hline
\multicolumn{1}{|c|}{\textbf{}}   & \multicolumn{1}{c|}{\textbf{\# atoms}} & \multicolumn{1}{c|}{\textbf{\# e-}} & \multicolumn{1}{c|}{\textbf{\begin{tabular}[c]{@{}c@{}}Density \\ fraction\end{tabular}}} & \multicolumn{1}{c|}{\textbf{\begin{tabular}[c]{@{}c@{}}Time/scf\\ (node-hrs)\end{tabular}}} & \multicolumn{1}{c|}{\textbf{\begin{tabular}[c]{@{}c@{}}ChF\\ (node-hrs)\end{tabular}}} & \multicolumn{1}{c|}{\textbf{\begin{tabular}[c]{@{}c@{}}Orth\\ (node-hrs)\end{tabular}}} & \multicolumn{1}{c|}{\textbf{\begin{tabular}[c]{@{}c@{}}Sub proj\\ (node-hrs)\end{tabular}}} & \multicolumn{1}{c|}{\textbf{\begin{tabular}[c]{@{}c@{}}Others\\ (node-hrs)\end{tabular}}} \\ \hline
\multicolumn{1}{|c|}{$\mathrm{Al_{13}}$}     & \multicolumn{1}{c|}{13}                & \multicolumn{1}{c|}{39}             & \multicolumn{1}{c|}{2.54e-03}                                                             & \multicolumn{1}{c|}{6.69e-04}                                                               & \multicolumn{1}{c|}{2.65e-04}                                                            & \multicolumn{1}{c|}{1.76e-04}                                                           & \multicolumn{1}{c|}{4.41e-05}                                                               & \multicolumn{1}{c|}{1.84e-04}                                                             \\ \hline
\multicolumn{1}{|c|}{$\mathrm{Al_{147}}$}     & \multicolumn{1}{c|}{147}               & \multicolumn{1}{c|}{441}            & \multicolumn{1}{c|}{3.86e-03}                                                             & \multicolumn{1}{c|}{6.86e-02}                                                               & \multicolumn{1}{c|}{4.67e-02}                                                            & \multicolumn{1}{c|}{3.89e-04}                                                           & \multicolumn{1}{c|}{6.53e-03}                                                               & \multicolumn{1}{c|}{1.50e-02}                                                             \\ \hline
\multicolumn{1}{|c|}{$\mathrm{Al_{561}}$}     & \multicolumn{1}{c|}{561}               & \multicolumn{1}{c|}{1683}           & \multicolumn{1}{c|}{7.57e-04}                                                             & \multicolumn{1}{c|}{0.693}                                                                  & \multicolumn{1}{c|}{0.443}                                                               & \multicolumn{1}{c|}{0.012}                                                              & \multicolumn{1}{c|}{0.052}                                                                  & \multicolumn{1}{c|}{0.186}                                                                \\ \hline
\multicolumn{1}{|c|}{$\mathrm{Al_{2057}}$}     & \multicolumn{1}{c|}{2057}              & \multicolumn{1}{c|}{6171}           & \multicolumn{1}{c|}{8.16e-04}                                                             & \multicolumn{1}{c|}{7.964}                                                                  & \multicolumn{1}{c|}{4.974}                                                               & \multicolumn{1}{c|}{0.461}                                                              & \multicolumn{1}{c|}{0.593}                                                                  & \multicolumn{1}{c|}{1.936}                                                                \\ \hline
\multicolumn{1}{|c|}{$\mathrm{Al_{6525}}$}    & \multicolumn{1}{c|}{6525}              & \multicolumn{1}{c|}{19575}          & \multicolumn{1}{c|}{6.33e-04}                                                             & \multicolumn{1}{c|}{55.077}                                                                 & \multicolumn{1}{c|}{32.606}                                                              & \multicolumn{1}{c|}{12.681}                                                             & \multicolumn{1}{c|}{4.108}                                                                  & \multicolumn{1}{c|}{5.682}                                                                \\ \hline
\multicolumn{9}{c}{(a)}                                                                                                                                                                                                                                                                                                                                                                                                                                                                                                                                                                                                                                                                   \\ \hline
\multicolumn{1}{|c|}{\textbf{}}   & \multicolumn{1}{c|}{\textbf{\# atoms}} & \multicolumn{1}{c|}{\textbf{\# e-}} & \multicolumn{1}{c|}{\textbf{\begin{tabular}[c]{@{}c@{}}Density \\ fraction\end{tabular}}} & \multicolumn{1}{c|}{\textbf{\begin{tabular}[c]{@{}c@{}}Time/scf\\ (node-hrs)\end{tabular}}} & \multicolumn{1}{c|}{\textbf{\begin{tabular}[c]{@{}c@{}}ChF\\ (node-hrs)\end{tabular}}} & \multicolumn{1}{c|}{\textbf{\begin{tabular}[c]{@{}c@{}}Orth\\ (node-hrs)\end{tabular}}} & \multicolumn{1}{c|}{\textbf{\begin{tabular}[c]{@{}c@{}}Sub proj\\ (node-hrs)\end{tabular}}} & \multicolumn{1}{c|}{\textbf{\begin{tabular}[c]{@{}c@{}}Others\\ (node-hrs)\end{tabular}}} \\ \hline
\multicolumn{1}{|c|}{$\mathrm{Si_{10}H_{16}}$}     & \multicolumn{1}{c|}{26}                & \multicolumn{1}{c|}{56}             & \multicolumn{1}{c|}{7.38e-03}                                                             & \multicolumn{1}{c|}{6.54e-03}                                                               & \multicolumn{1}{c|}{1.67e-03}                                                            & \multicolumn{1}{c|}{7.88e-04}                                                           & \multicolumn{1}{c|}{2.04e-04}                                                               & \multicolumn{1}{c|}{3.88e-03}                                                             \\ \hline
\multicolumn{1}{|c|}{$\mathrm{Si_{220}H_{144}}$}   & \multicolumn{1}{c|}{364}               & \multicolumn{1}{c|}{1024}           & \multicolumn{1}{c|}{1.83e-04}                                                             & \multicolumn{1}{c|}{9.37e-02}                                                               & \multicolumn{1}{c|}{5.94e-02}                                                            & \multicolumn{1}{c|}{2.26e-03}                                                           & \multicolumn{1}{c|}{7.11e-03}                                                               & \multicolumn{1}{c|}{2.49e-02}                                                             \\ \hline
\multicolumn{1}{|c|}{$\mathrm{Si_{525}H_{276}}$}   & \multicolumn{1}{c|}{801}               & \multicolumn{1}{c|}{2376}           & \multicolumn{1}{c|}{6.58e-04}                                                             & \multicolumn{1}{c|}{0.964}                                                                  & \multicolumn{1}{c|}{0.609}                                                               & \multicolumn{1}{c|}{0.033}                                                              & \multicolumn{1}{c|}{0.081}                                                                  & \multicolumn{1}{c|}{0.241}                                                                \\ \hline
\multicolumn{1}{|c|}{$\mathrm{Si_{1214}H_{504}}$}  & \multicolumn{1}{c|}{1718}              & \multicolumn{1}{c|}{5360}           & \multicolumn{1}{c|}{4.16e-04}                                                             & \multicolumn{1}{c|}{3.853}                                                                  & \multicolumn{1}{c|}{2.674}                                                               & \multicolumn{1}{c|}{0.292}                                                              & \multicolumn{1}{c|}{0.319}                                                                  & \multicolumn{1}{c|}{0.568}                                                                \\ \hline
\multicolumn{1}{|c|}{$\mathrm{Si_{6047}H_{1308}}$} & \multicolumn{1}{c|}{7355}              & \multicolumn{1}{c|}{25496}          & \multicolumn{1}{c|}{7.38e-04}                                                             & \multicolumn{1}{c|}{67.492}                                                                 & \multicolumn{1}{c|}{43.211}                                                              & \multicolumn{1}{c|}{13.800}                                                             & \multicolumn{1}{c|}{5.579}                                                                  & \multicolumn{1}{c|}{4.902}                                                                \\ \hline
\multicolumn{9}{c}{(b)}                                                                                                                                                                                                                                                                                                                                                                                                                                                                                                                                                                                                                                                                  
\end{tabular}
\caption{ Breakdown of computational times for the various steps in the solution of the Kohn-Sham equations in the localized Tucker tensor basis using the Chebyshev filtering based subspace iteration. The benchmark systems considered are: (a) aluminum nano-particles; (b) silicon quantum dots.
}
\label{tab:time breakdown}
\end{table}

\end{document}